\documentclass[nature
,showpacs
,superscriptaddress
]
{revtex4-2}
\usepackage{newfloat,algcompatible}
\usepackage{etoolbox}
\AtBeginEnvironment{algorithm}{\noindent\hrulefill\par\nobreak\vskip-5pt}
\DeclareFloatingEnvironment[
    fileext=loa,
    listname=List of Algorithms,
    name=ALGORITHM,
    placement=tbhp,
]{algorithm}
\usepackage{multirow}
\usepackage{graphicx,subfigure,bbm}

\usepackage{amsmath}
\usepackage{tabularx}
\usepackage{amsthm}
\usepackage{amssymb}
\usepackage{array,bm}
\usepackage{graphicx}
\usepackage{fancyhdr}
\usepackage{color}
\usepackage{extarrows}
\usepackage{enumerate}
\usepackage{epstopdf}
\usepackage[colorlinks,
            linkcolor=black,
            citecolor=black,
            urlcolor=blue
            ]{hyperref}
\usepackage{natbib}
\usepackage[latin1]{inputenc}
\usepackage{tikz}
\usetikzlibrary{shapes,arrows}
\usepackage{booktabs}
\usepackage{subfigure}
\usepackage{parskip}
\usepackage{lineno}
\usepackage{lipsum}

\tikzstyle{decision} = [diamond, draw, fill=blue!20,
    text width=4.5em, text badly centered, node distance=3cm, inner sep=0pt]
\tikzstyle{block} = [rectangle, draw, fill=blue!20,
    text width=10em, text centered, rounded corners, minimum height=4em]
\tikzstyle{line} = [draw, -latex']
\tikzstyle{cloud} = [rectangle, draw,fill=red!20, node distance=7cm,
    minimum height=4em]

\def\(({\left(}
\def\)){\right)}
\def\[[{\left[}
\def\]]{\right]}

\newcommand{\be}{\begin{equation}}
\newcommand{\ee}{\end{equation}}
\newcommand{\bea}{\begin{eqnarray}}
\newcommand{\eea}{\end{eqnarray}}

\DeclareMathAlphabet{\varmathbb}{U}{bbold}{m}{n}

\begin{document}

\title{Neural-network solutions to stochastic reaction networks}

\author{Ying Tang}
\email[These authors contributed equally; Corresponding authors: ]{jamestang23@gmail.com}
\affiliation{International Academic Center of Complex Systems, Beijing Normal University, Zhuhai 519087, China}

\author{Jiayu Weng}
\email[These authors contributed equally]{}
\affiliation{Faculty of Arts and Sciences, Beijing Normal University, Zhuhai 519087, China}
\affiliation{School of Systems Science, Beijing Normal University, Beijing 100875, China}

\author{Pan Zhang}
\email[Corresponding authors: ]{panzhang@itp.ac.cn}
\affiliation{CAS Key Laboratory for Theoretical Physics, Institute of Theoretical Physics, Chinese Academy of Sciences, Beijing 100190, China}
\affiliation{School of Fundamental Physics and Mathematical Sciences, Hangzhou Institute for Advanced Study, UCAS, Hangzhou 310024, China}
\affiliation{International Centre for Theoretical Physics Asia-Pacific, Beijing/Hangzhou, China}

\begin{abstract}
The stochastic reaction network in which chemical species evolve through a set of reactions is widely used to model stochastic processes in physics, chemistry and biology. To characterize the evolving joint probability distribution in the state space of species counts requires solving a system of ordinary differential equations, the chemical master equation, where the size of the counting state space increases exponentially with the type of species, making it challenging to investigate the stochastic reaction network. Here, we propose a machine-learning approach using the variational autoregressive network to solve the chemical master equation. Training the autoregressive network employs the policy gradient algorithm in the reinforcement learning framework, which does not require any data simulated in prior by another method. Different from simulating single trajectories, the approach tracks the time evolution of the joint probability distribution, and supports direct sampling of configurations and computing their normalized joint probabilities. We apply the approach to representative examples in physics and biology, and demonstrate that it accurately generates the probability distribution over time. The variational autoregressive network exhibits a plasticity in representing the multimodal distribution, cooperates with the conservation law, enables time-dependent reaction rates, and is efficient for high-dimensional reaction networks with allowing a flexible upper count limit. The results suggest a general approach to investigate stochastic reaction networks based on modern machine learning.
\end{abstract}

\maketitle


The stochastic reaction network is a standard model for stochastic processes in  physics \cite{weber2017master}, chemistry \cite{gillespie2007stochastic,ge2012stochastic}, biology \cite{elowitz2002stochastic}, and ecology \cite{blythe2007stochastic,PhysRevLett.115.158101}. Representative examples include the birth-death process \cite{gardiner2004handbook}, the model of spontaneous asymmetric synthesis \cite{frank1953spontaneous}, and gene regulatory networks \cite{bressloff2014stochastic}. In particular, due to the rapid development of measuring molecules at the single-cell level, it becomes increasingly important to model intracellular reaction networks which  have low counts of molecules and are subject to random noise \cite{raj2008nature}. These stochastic reaction networks in a well-mixed condition can be modeled by the chemical master equation (CME)  \cite{van2007stochastic}, which describes a time-evolving joint probability distribution of discrete states representing counts of reactive species. However, the number of possible states grows exponentially with the number (type) of species; thus, it is 
challenging to exactly represent the joint probability and solve the CME. 

Many efforts have been made to approximately solve the CME by numerical methods. The most popular method, the Gillespie algorithm \cite{doob1942topics,gillespie1976general,weinan2021applied} as a type of the kinetic Monte Carlo method,  
samples from all possible trajectories to generate statistics of relevant variables. However, achieving high-accuracy statistics of the joint probability distribution requires a large number of trajectories. Moreover, the dynamics can be dramatically affected by rare but important trajectories, which are difficult to sample by the Gillespie algorithm \cite{terebus2019discrete,terebus2021exact}. Different from the sampling-based methods, asymptotic approximations have been proposed to transform the CME into continuous-state equations, e.g., the chemical Langevin equation \cite{gillespie2000chemical}. This method is more computationally efficient, 
but the continuous-state approximation becomes inaccurate when fluctuations on the count of species are significant, e.g., in the case of proteins  \cite{raj2008nature}. 
Another class of methods truncates the CME into a state space covering the majority of the probability distribution, including the finite state projection \cite{munsky2006finite}, the sliding window method \cite{henzinger2009sliding}, and the ACME method \cite{cao2016accurate,cao2016state}. 
Further advances employ the Krylov subspace approximation \cite{doi:10.1137/060678154} and tensor-train representations \cite{kazeev2014direct,ion2021tensor}. However, the computational cost of these methods is still prohibitive to reach high accuracy when both the number  and counts of species become large~\cite{gupta2021deepcme}.
Although a great effort has been made, we still lack a general method to solve the CME by directly facing the representation problem of the evolving joint probability distribution. 

\begin{figure}[ht]
{\includegraphics[width=0.96\textwidth]{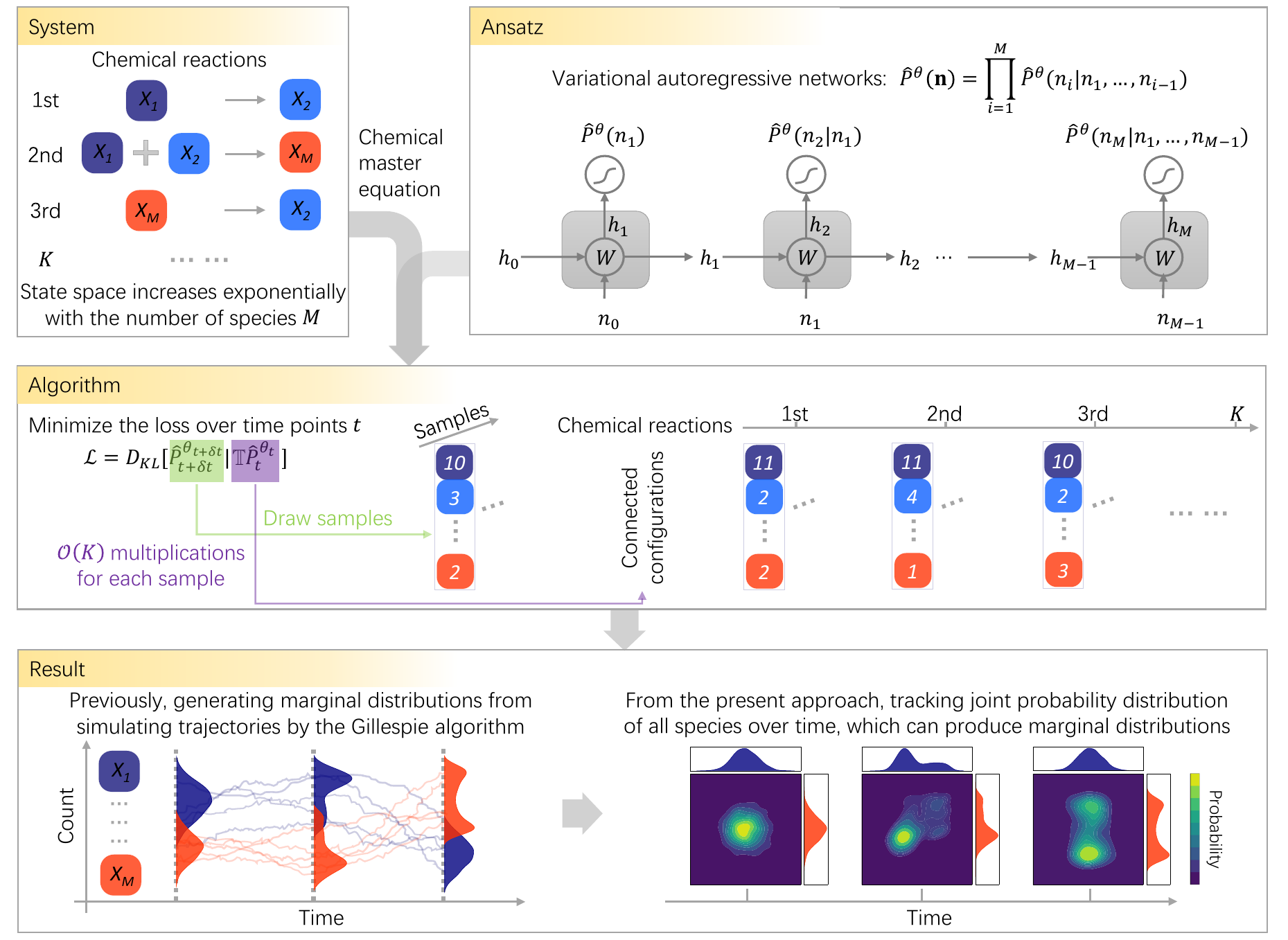}}
\caption{Tracking the joint probability distribution of stochastic reaction networks over time. (Upper) For a reaction network, the state space scales exponentially with the number of species $M$, making it difficult to track the time evolution of the joint distribution. 
The ansatz VAN, such as with the unit of the RNN, can represent the joint distribution 
$\hat{P}^{\theta}(\textbf{n})$. (Middle) Starting from an initial distribution, we minimize the loss function by the KL-divergence between joint distributions at consecutive time steps to learn its time evolution. The subscript $t$ denotes time points, $\theta$ denotes parameters of the neural network to be optimized, and $\mathbb{T}$ is the transition kernel of the CME. To train the VAN at time $t+\delta t$, samples are drawn from the distribution $\hat{P}^{\theta_{t+\delta t}}_{t+\delta t}$. Each sample is illustrated by a column of stacked squares with color specifying the species and their counts inside the square. For each sample, the number of connected configurations, e.g., those transit into the sample through the transition operator, is equal to the number of chemical reactions $K$. 
(Bottom) The previous method of simulating trajectories by the Gillespie algorithm can generate marginal distributions, but is computationally prohibitive to accurately produce high-dimensional joint distributions. Here, the VAN tracks the joint distribution of all species over time. 
}
\label{Fig1}
\end{figure}

In this paper, we develop a neural-network approach to investigate the time evolution of the joint probability distribution for the stochastic reaction network. 
Our method is inspired by the strong representation power  
of neural networks for high-dimensional data \cite{mehta2019high,RevModPhys.91.045002,RevModPhys.91.045002,Tang_2022}. In particular, we leverage the variational autoregressive network (VAN)  to solve the CME (Fig.~\ref{Fig1}). The VAN has been applied to statistical physics \cite{PhysRevLett.122.080602}, quantum many-body systems \cite{PhysRevResearch.2.023358,PhysRevLett.124.020503,barrett2022autoregressive}, open quantum systems \cite{PhysRevLett.128.090501}, quantum circuits \cite{PhysRevA.104.032610} and computational biology \cite{shin2021protein}, where it enables to efficiently sample configurations and compute the normalized probability of configurations. 
Here, we extend the VAN to characterize the joint probability distribution of species counts for the stochastic reaction network. As the unit of the VAN, we have employed recurrent neural networks (RNN) \cite{PhysRevResearch.2.023358} and the transformer \cite{NIPS2017_3f5ee243}, which are flexible to represent the high-dimensional probability distribution and to adjust the upper count limit. We also enable the VAN to have the count constraint on each species 
or to maintain the conservation on the total count of species for specific systems, both of which can improve the accuracy with the contracted probability space.

The present approach differs significantly from the recently developed methods. Instead of using the simulated data from the Gillespie algorithm to train the neural network \cite{jiang2021neural,sukys2022approximating}, our approach does not need the aid of any existing data  or measurements. This is advantageous, especially for the cases where the Gillespie algorithm itself is inefficient to capture the multimodal distribution \cite{terebus2021exact}. 
The present approach gives an automatically normalized distribution as the solution of the CME at arbitrary finite time, different from learning the transition kernels \cite{bortolussi2018deep}. 
The obtained joint distribution contains more information than estimating the marginal statistics alone \cite{gupta2021deepcme}, providing the probability for each configuration in the high-dimensional state space.

To demonstrate the advances of the proposed approach, we apply it to representative examples of stochastic reaction networks, 
including the genetic toggle switch with multimodal distribution \cite{terebus2019discrete}, the intracellular signaling cascade in the high dimension \cite{gupta2021deepcme}, the early life self-replicator with an intrinsic constraint of count conservation \cite{PhysRevLett.115.158101}, and the epidemic model with time-dependent rates \cite{thanh2015simulation}. 
The results on the marginal statistics match those from the previous numerical methods, such as the Gillespie algorithm or the finite state projection \cite{munsky2006finite}. The present approach further efficiently produces the time evolution of the joint probability distribution. 
Especially, it can learn the multimodal distribution, 
is effective for systems with feedback regulations, and is computationally efficient for the high-dimensional system, where the computational time scales almost linearly to the number of species, 
allowing the approach to be applicable and adaptable to general stochastic reaction networks.

\subsection{Chemical master equation}

We consider the discrete-state continuous-time Markovian dynamics with configurations $\textbf{n}=\{n_{1},n_{2},\dots,n_{M}\}$ and the size of variables $M$.
The probability distribution at time $t$ evolves under the stochastic master equation  \cite{gardiner2004handbook}:
\begin{align}
\label{ME}
\partial_{t}P_{t}(\textbf{n})=\sum_{\textbf{n}^{'}\neq\textbf{n}}W_{\textbf{n}^{'}\rightarrow\textbf{n}}P_{t}(\textbf{n}^{'})-r_{\textbf{n}}P_{t}(\textbf{n}),
\end{align}
where $P_{t}(\textbf{n})$ is the probability of the configuration state $\textbf{n}$, $W_{\textbf{n}\rightarrow\textbf{n}^{'}}$ are the transition rates from $\textbf{n}$ to $\textbf{n}^{'}$, and the escape rate from $\textbf{n}$ is $r_{\textbf{n}}=\sum_{\textbf{n}^{'}\neq\textbf{n}}W_{\textbf{n}\rightarrow\textbf{n}^{'}}$. As the probabilities of transiting into and out from any configuration sum up to $0$, the total probability is conserved over time.

One type of the master equation describing stochastic reaction networks is the CME. 
Specifically, a stochastic reaction network with $K$ reactions and $M$ species (each species $X_{i}$ has the count $n_{i}=0,1,\dots N_{i}$ with $1\leq i\leq M$) is:
\begin{align}
\sum_{i=1}^{M}r_{ki}X_{i}\xrightarrow{c_{k}}\sum_{i=1}^{M}p_{ki}X_{i},
\end{align}
with the reaction rates $c_{k}$ for the $k$-th reaction. The numbers of reactants and products are denoted by $r_{ki}$ and $p_{ki}$, respectively. 
The change in the count of species follows the stoichiometric matrix: $s_{ki}=p_{ki}-r_{ki}$. To formulate the CME, a set of propensity functions need to be specified. The propensities, $a_{k}(\textbf{n})$, represent the probability of the reaction $k$ occurring in the infinitesimal time interval at the state $\textbf{n}$ \cite{gillespie2007stochastic}. For example, a conventional way follows the law of mass action \cite{bressloff2014stochastic}: the propensity $a_{k}$ for each reaction is calculated by multiplying the reaction rate $c_{k}$ and the count of species $n_{i}$ to the power of $r_{ki}$. 

Given the stoichiometric matrix, reaction rates, propensities and an initial distribution $P_{0}(\textbf{n})$, the system evolves according to the CME \cite{van2007stochastic}:
\begin{align}
\label{CME}
\partial_{t}P_{t}(\textbf{n})  = \sum_{k=1}^{K} [a_{k}(\textbf{n}-s_{k})P_{t}(\textbf{n}-s_{k}) - a_{k}(\textbf{n})P_{t}(\textbf{n})],
\end{align}
where $s_{k}$ is the $k$-th row of the stoichiometric matrix. We consider the reflecting boundary condition for both boundaries at $n_{i}=0$ and $n_{i}=N_{i}$, by setting the transition probability out of the boundary to zero, where the CME conserves the probability over time. Other boundary conditions 
can be employed when necessary. 
The state space of the probability distribution scales exponentially with the number of species: $N^{M}$ if each species has a count up to $N$. The exponentially increasing state space makes it challenging to solve the CME. 
Based on a neural-network ansatz, we next provide a computational approach toward solving this problem.

\section{Results}

\subsection{Design of the VAN}

We leverage the VAN \cite{PhysRevLett.122.080602} to represent the probability distribution of species counts in the CME. The VAN is a product of conditional probabilities:
\begin{align}
\label{VAN}
\hat{P}^{\theta}(\textbf{n})=\prod^{M}_{i=1}\hat{P}^{\theta}(n_{i}|n_{1},\dots,n_{i-1}),
\end{align}
where $n_{i}$ is the count of the $i$-th species, and $\theta$ denotes the parameters of the VAN to be trained.  The VAN parameterizes an automatically normalized  probability distribution, such that sampling configurations can be performed in parallel under the corresponding probabilities.  Compared with the VAN for statistical mechanics \cite{PhysRevLett.122.080602} where each variable has binary states for the spin, we extend it to the $N$-state variable, where $N$ denotes the maximum of the upper count limit of all the species: $N=\max_{i}N_{i}$. 
The issue of choosing a proper $N$ is discussed in Methods. 

The conventional unit of the VAN includes MADE \cite{MathieuGermain2015MADEMA}, PixelCNN \cite{van2016pixel}, RNN \cite{PhysRevResearch.2.023358} and transformer \cite{NIPS2017_3f5ee243}. We mainly employ the RNN and transformer to represent the highly correlated distributions (Methods). In the examples, we find that a single-layer RNN is sufficient to parameterize the joint distribution, and the transformer performs equally well. 
For specific systems, there may be a priori knowledge on the count of certain species or an intrinsic conservation on the total count of all species. These constraints can be put into the VAN to improve the accuracy of training in the contracted probability space (Methods). 

\subsection{Tracking time evolution of the probability distribution}

To track the probability distribution over time, we minimize the loss function:
\begin{align}
\label{Loss}
\mathcal{L}&= D_{KL}[\hat{P}^{\theta_{t+\delta t}}_{t+\delta t}|\mathbb{T}\hat{P}^{\theta_{t}}_{t}],
\end{align}
which is the Kullback-Leibler (KL)-divergence between a distribution $\hat{P}^{\theta_{t+\delta t}}_{t+\delta t}$ parameterized by the VAN and the one-time-step 
evolution of the distribution $\hat{P}^{\theta_{t}}_{t}$ at time $t$. The transition kernel in the time interval $\delta t$ is $\mathbb{T}= e^{\delta t \mathbb{W}}\approx(I+\delta t \mathbb{W})$, which preserves the total probability, and the generator $\mathbb{W}$ is given by Eq.~\eqref{ME}. 
The time-step length $\delta t$ is typically fixed, but an adaptive $\delta t$ can be used to help accelerate the simulation (Methods).

The loss function is approximated by drawing samples from the VAN:
\begin{align}
\label{loss2}
\mathcal{L}&=
\mathbb{E}_{\textbf{s}\sim\hat{P}^{\theta_{t+\delta t}}_{t+\delta t}}\{\ln \hat{P}^{\theta_{t+\delta t}}_{t+\delta t}(\textbf{s})-\ln[(\mathbb{T}\hat{P}^{\theta_{t}}_{t})(\textbf{s})]\},
\end{align}
where $\textbf{s}$ are samples drawn from the distribution $\hat{P}^{\theta_{t+\delta t}}_{t+\delta t}$ parameterized by the VAN. At each time step, we calculate the manipulation of the configurations transiting in and out of each drawn sample, which are also paralleled for all the samples. The number of such ``connected'' configurations for each sample in the CME is equal to the number of chemical reactions $K$. Thus, only the order of $\mathcal{O}(K)$ multiplications is required to account for all the possible transitions of each configuration, rather than performing $\mathcal{O}(N^M)$ multiplications of the full transition matrix. 
This crucial property leads to an efficient method to approximately solve the CME by the VAN.

\begin{algorithm*}
\label{algorithm1}
\begin{algorithmic}
\\\hrulefill
\State{\textbf{Input}: System dimension, the stoichiometric matrix, reaction rates,  
propensities, the initial distribution,  hyperparameters (the upper count limit, time steps, step length, size of the VAN, learning rate, batch size, the number of epochs).}
\State{\textbf{Output}: The joint probability distributions over time and relevant statistics.}
\\\hrulefill
\State{Initialize the distribution, choose hyperparameters of the VAN.}
\FOR{Every time step}
       \STATE{Learn the next-step VAN $\hat{P}^{\theta_{t+\delta t}}_{t+\delta t}$:}
        \FOR{Every epoch}
        \STATE{1. Draw samples from VAN, $\textbf{s}\sim\hat{P}^{\theta_{t+\delta t}}_{t+\delta t}$. }
         \STATE{2. Generate connected configurations, which transit from or into the sample $\textbf{s}$ through the chemical reactions.}
         \STATE{3. Acquire the probabilities of the connected configurations from $\hat{P}^{\theta_{t}}_{t}$.}
        \STATE{4. Calculate the matrix elements of the transition kernel $\mathbb{T}$ for the connected configurations.}
        \STATE{5. Train the VAN $\hat{P}^{\theta_{t+\delta t}}_{t+\delta t}$ by minimizing the loss function:
        $\mathcal{L}=\mathbb{E}_{\textbf{s}\sim\hat{P}^{\theta_{t+\delta t}}_{t+\delta t}}
        \{\ln \hat{P}^{\theta_{t+\delta t}}_{t+\delta t}(\textbf{s})-\ln[(\mathbb{T}\hat{P}^{\theta_{t}}_{t})(\textbf{s})]\}$.}
         \STATE{6. Save the VAN $\hat{P}^{\theta_{t+\delta t}}_{t+\delta t}$ for the next time step.}
        \ENDFOR
        \STATE{Calculate statistics from the joint distribution $\hat{P}^{\theta_{t+\delta t}}_{t+\delta t}$.} 
\ENDFOR
\STATE{Search for optimal hyperparameters, by evaluating the loss function or comparing with other methods (optional).} 
\\\hrulefill
\end{algorithmic}
\caption{Solving the CME by the VAN. 
The choice of hyperparameters can be based on  Table~\uppercase\expandafter{\romannumeral1}.}
\end{algorithm*}

The VAN is updated by:
\begin{align}
\label{Training}
\nabla_{\theta_{t+\delta t}}\mathcal{L}&=\mathbb{E}_{\textbf{s}\sim\hat{P}^{\theta_{t+\delta t}}_{t+\delta t}}\{[\nabla_{\theta_{t+\delta t}}\ln \hat{P}^{\theta_{t+\delta t}}_{t+\delta t}(\textbf{s})]\cdot\{\ln \hat{P}^{\theta_{t+\delta t}}_{t+\delta t}(\textbf{s})-\ln[(\mathbb{T}\hat{P}^{\theta_{t}}_{t})(\textbf{s})]\}\},
\end{align}
where we have used $\mathbb{E}_{\textbf{s}\sim\hat{P}^{\theta_{t+\delta t}}_{t+\delta t}}[\nabla_{\theta_{t+\delta t}}\ln \hat{P}^{\theta_{t+\delta t}}_{t+\delta t}(\textbf{s})]=\nabla_{\theta_{t+\delta t}}\sum_{\textbf{s}}\hat{P}^{\theta_{t+\delta t}}_{t+\delta t}(\textbf{s})=\nabla_{\theta_{t+\delta t}}1=0$ when sufficient samples are drawn. 
Training by the gradient Eq.~\eqref{Training} amounts to the policy gradient algorithm in the reinforcement learning framework \cite{williams1992simple}: the term in the curly bracket of Eq.~\eqref{loss2} is the reward, and  $\hat{P}^{\theta_{t+\delta t}}_{t+\delta t}(\textbf{s})$ is a stochastic policy.  
The pseudocode is summarized in ALGORITHM~1.


\subsection{Examples}

We apply our approach to representative examples of stochastic reaction networks, including the genetic toggle switch \cite{terebus2019discrete} and the intracellular signaling cascade \cite{gupta2021deepcme}. 
They separately demonstrate that our approach is applicable to systems with a multimodal distribution 
and in high dimensions. 
More examples are provided in the Supplementary: the birth-death process (Supplementary Fig.~2), gene expression without regulation (Supplementary Fig.~3),  an autoregulatory feedback loop \cite{sukys2022approximating} (Supplementary Fig.~4), the early life self-replicator  \cite{PhysRevLett.115.158101} (Supplementary Fig.~5), the epidemic model \cite{thanh2015simulation} (Supplementary Fig.~6), where the last two examples respectively have an intrinsic constraint of count conservation or time-dependent parameters.  
The computational time and the hyperparameters of the VAN for all the examples are in Table~\ref{table1}. With the VAN, tracking the joint distribution becomes computationally efficient for stochastic reaction networks. 

To benchmark our method, we compare with 
the other numerical methods.  
Specifically, the Gillespie simulation is used as the comparison of the marginal statistics for all the examples.   
The finite state projection agrees with our result in the autoregulatory feedback loop with the bimodal distribution of the protein. 
We also find a match between the VAN and the ACME method  \cite{terebus2019discrete} on learning the multimodal distribution for the genetic toggle switch. 
To highlight the effectiveness of the VAN on capturing the joint distribution, in each example we show the joint distribution for two of the species (heatmaps). We next provide the results for the  toggle switch  and the intracellular signaling cascade. 

\begin{table}[ht]
\begin{tabular}{ |c|c|c|c|c|c|c|c|c|c|c|  }
\hline
  &Species  &  Reactions  &  $N$ &Time steps &$\delta t$ & Epochs & Depth & Width  &Comput. time (hr) \\
\hline
Birth-death process & 1 & 2 & 30 & $10^{4}$ & $10^{-2}$ & 100 &  1 & 16 &1.62 \\
\hline
Gene expression & 2 & 4 & 100&   $3.6*10^{4}$ & $10^{-1}$ &100 & 1 & 32 & 10.21 \\
\hline
Autoregulatory feedback & 2 & 5& 100  & $2*10^{3}$ & $5*10^{-3}$ & 100 & 1 & 32  &0.65 \\
\hline
Early life self-replicator & 3 & 6& 100  & $2*10^{5}$ & $4*10^{-4}$ & 20 & 1 & 32  & 42.21 \\
\hline
Epidemic model & 3 & 6& 80  & $10^{4}$ & $2*10^{-2}$ & 100 & 1 & 8  & 3.68 \\
\hline
Toggle switch & 4 & 8 & 80 & $8*10^{3}$ & $5*10^{-3}$ & 100 & 1 & 32  & 4.23 \\
\hline
Signaling cascade $1$ & 15&30 & 10   & $10^{3}$ & $10^{-2}$& 100 &  1 & 32  &1.66\\
\hline
Signaling cascade $2$& 15& 30&  40  & $10^{4}$ & $10^{-3}$ & 100 &  1 & 32  &16.68\\
\hline
Signaling cascade $3$ & 15 & 31 & 10 & $10^{3}$ & $10^{-2}$ & 100 & 1 & 32  &1.66 \\
\hline
\end{tabular}
\caption{The models solved by the present approach and the computational time under the chosen hyperparameters. The  time-step length $\delta t$ has the unit of the inverse of the reaction rates. 
The computational time for various numbers of species in the  signaling cascade $1$ is shown in  Fig.~\ref{Fig6}f. 
The number of epochs is that used at each time step, except for the first time step requiring $\mathcal{O}(10^{3})$ epochs to converge, with training details in Methods. The depth and width are for the RNN, whereas the results from the transformer for the signaling cascade are in Supplementary Table~\uppercase\expandafter{\romannumeral2}. 
All computations are performed with a single core GPU ($\sim25\%$ usage) of a Tesla-V100. }
\label{table1}
\end{table}


\subsubsection{Genetic toggle switch}

\begin{figure}[htbp]
{\includegraphics[width=1\textwidth]{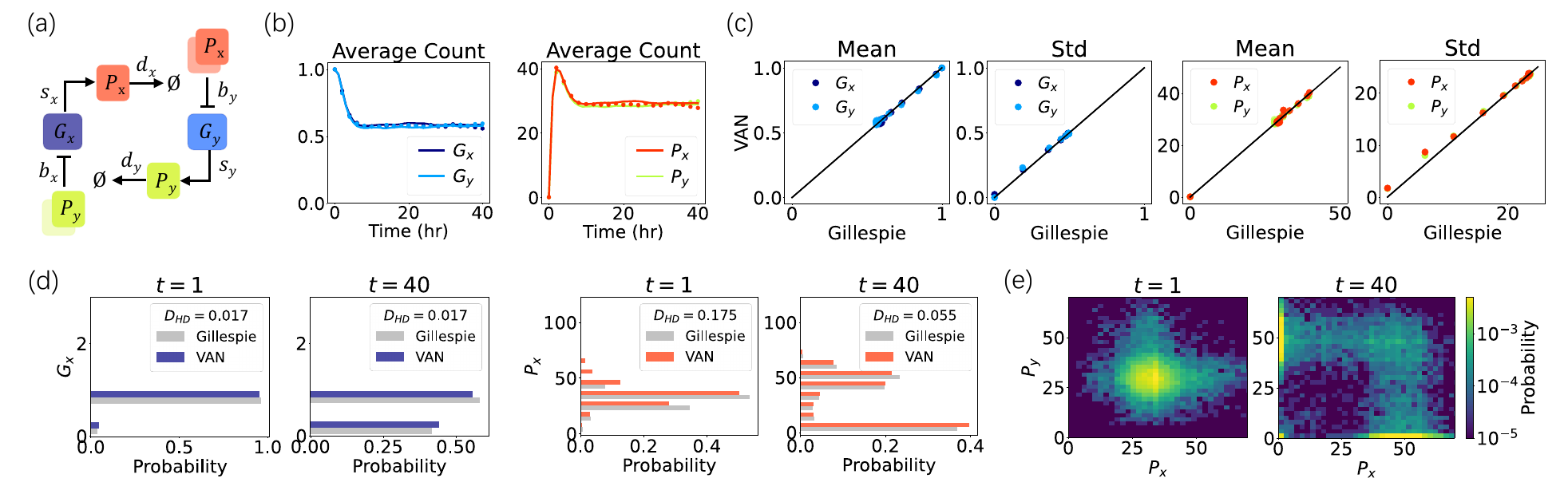}}
\centering
\caption{The result of the genetic toggle switch. (a) A schematic of the reactions. (b) The time evolution of the average count for the genes and proteins,  from the VAN (dots) and the Gillespie simulation (lines). 
(c) Comparison of the means and standard deviations of genes and proteins between the VAN and the Gillespie simulation, at time points $t = 0, 1, ..., 40$. (d) The marginal  distributions of the gene $G_{x}$ and protein $P_{x}$ at time points $t = 1, 40$ from the Gillespie simulation (gray) and the VAN. 
The inset is the Hellinger distance between the two distributions.
(e) The joint distribution of the two proteins from the VAN at time points $t=1, 40$, with the color bar for the probability values in the logarithmic scale. In the long-time limit, there are  four stable states with probability peaks located at $(P_{x}, P_{y}) =\{ (0, 0), (50, 0), (0, 50), (50, 50)\}$. 
The Gillespie simulation has $10^{4}$ trajectories. 
The initial distribution is the delta distribution with one unbounded form for each gene and zero proteins. The hyperparameters of the VAN are in Table~\uppercase\expandafter{\romannumeral1}, and the parameters of the system are given in Supplementary Note~2. 
}
    \label{Fig3}
\end{figure}

In the toggle switch \cite{gardner2000construction,terebus2019discrete} (Fig.~\ref{Fig3}a), there are six molecular species: genes $G_{x}$, $G_{y}$, which transcript two proteins $P_{x}$, $P_{y}$ inhibiting the gene expression of each other, and two protein-DNA complexes $\bar{G}_{x}$ ($\bar{G}_{y}$) with protein $P_{y}$ ($P_{x}$) bound to gene $G_{x}$ ($G_{y}$). The dimer of the protein $P_{x}$ ($P_{y}$) inhibits the activity of the gene $G_{y}$ ($G_{x}$). The result from the VAN  matches those from the Gillespie simulation (Fig.~\ref{Fig3}, Supplementary Fig.~7), including the mean  (Fig.~\ref{Fig3}b) and standard deviation (Fig.~\ref{Fig3}c). The marginal distributions (Fig.~\ref{Fig3}d) on the count of the gene $G_{x}$ and protein $P_{y}$ are also consistent between the two methods. The joint distribution of the two proteins (Fig.~\ref{Fig3}e) shows the multimodality corresponding to the four stable genetic states of the two genes:  ``Off-Off'' ($G_{x}=0, G_{y}=0$), ``On-Off'' ($G_{x}=1, G_{y}=0$), ``Off-On'' ($G_{x}=0, G_{y}=1$), and ``On-On'' ($G_{x}=0, G_{y}=0$), which is consistent with the 
ACME method shown in Fig.~4 of  \cite{terebus2019discrete}. 

\subsubsection{Intracellular signaling cascade}

\begin{figure}[ht]
{\includegraphics[width=0.96\textwidth]{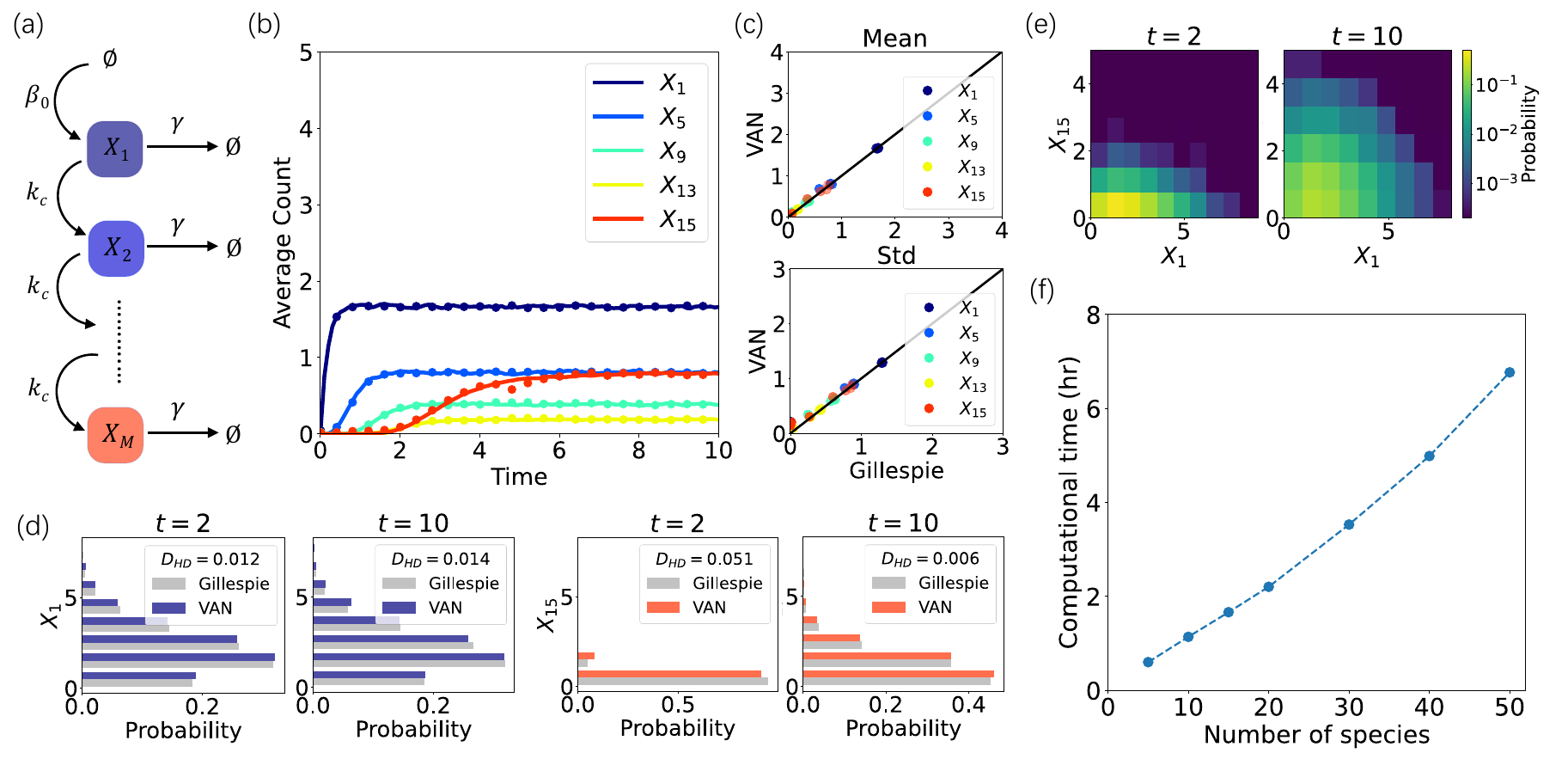}}
\caption{The result of the linear signaling cascade. (a) A schematic of the chemical reaction. (b) The time series of the average count of species from the VAN (dots) and the Gillespie simulation (lines). The color specifies the chemical species. 
(c) Comparison of the means and standard deviations of the counts between the VAN and the Gillespie simulation, at time points $t=1, 2, ..., 10$. 
(d) The marginal count distributions of various species are plotted horizontally at time points $t=2, 10$. 
The inset contains the Hellinger distance between the two distributions. 
(e) The joint distribution of $X_{1}$ and $X_{15}$ at time points $t=2, 10$ from the VAN, with the color bar denoting the probability values. 
The initial distribution is the delta distribution with all species zero, and the Gillespie simulation has $10^{4}$ trajectories. The parameters are $M=15$, $\beta_{0}=10$,  $k_{c}=5$, $\gamma=1$.  
(f) The computational time for the system 
with various number of species, under $1000$ time steps and step length $\delta t=0.01$. The hyperparameters of the VAN are in Table~\uppercase\expandafter{\romannumeral1}.
}
\label{Fig6}
\end{figure}

The intracellular signaling cascade \cite{gupta2021deepcme} is a biological reaction with a series of reactions, where one species catalyzes the production of the next (Fig.~\ref{Fig6}a). This recurrent structure of the signaling cascade makes it an ideal model to conveniently test our method with different numbers of species. 
The time evolution of the mean species count (Fig.~\ref{Fig6}b) and standard deviations (Fig.~\ref{Fig6}c) from the VAN match with the Gillespie simulation.  
The marginal distributions  from the two methods (Fig.~\ref{Fig6}d) also has a small Hellinger distance. 
Besides generating marginal statistics, the VAN gives a joint probability distribution of all the species (Fig.~\ref{Fig6}e).  
The computational time scales almost linearly to the number of species (Fig.~\ref{Fig6}f), with the result of $M=50$ species in Supplementary Fig.~8. 
The accuracy of the VAN is not sensitive to the order of the species used for the conditional probabilities in Eq.~\eqref{VAN} (Supplementary Fig.~9). 
The transformer has a similar accuracy compared with the RNN (Supplementary Fig.~10), but requires longer computational time. 
The signaling cascade with the nonlinear activation or with feedback regulations is accurately tracked by the VAN as well (Supplementary Figs.~11,~12). 

\section{Discussion}

We have developed a machine-learning approach to track stochastic reaction networks, by leveraging the VAN for the joint probability distribution in the CME. Specifically, there can be hundreds of possible states for each species, beyond the binary spin state in statistical physics  \cite{PhysRevLett.122.080602}. During the time evolution, the connected configurations of each sample depend on the chemical reactions. 
These properties require a tailored architecture of the VAN and rule of the time evolution. We have achieved such an extension, leading to the first approach of using the neural network alone to solve the CME. 
We have applied it to examples in biophysics, systems biology and epidemics, where it is accurate to track the joint probability distribution over time and is efficient in high-dimensional systems.


Learning the multimodal probability distribution can be affected by the mode-collapse \cite{PhysRevLett.122.080602}, since the variational training employs the reverse KL-divergence. Still, the VAN has effectively learned the bimodal distribution in the example of the early life self-replicator and autoregulatory feedback loop, and captured the multimodality in the toggle switch \cite{terebus2019discrete}. 
This is due to that the VAN continuously track the evolving distribution from the initial delta distribution with a single peak, such that it can capture the gradual change of the distribution when new probability peaks progressively emerge, similar to the annealed importance sampling \cite{neal2001annealed}. 
To alleviate the mode-collapse in complex problems with rugged distributions, the temperature annealing \cite{PhysRevLett.122.080602} and variational annealing \cite{hibat2021variational} are also helpful. 
Furthermore, sampling rare configurations may be necessary to capture the probability peaks with relatively lower probabilities, for which an importance sampling can be employed \cite{tang2022solving}. 


With the joint probability distribution learnt by the VAN, evaluating the macroscopic thermodynamical quantities including the free energy and entropy is achievable \cite{PhysRevLett.122.080602}. 
The present approach may also be adapted to reaction networks in more complex scenarios, such as non-Markovian dynamics \cite{jiang2021neural}. Beyond the well-mixed condition, the spatially-extended models can have hundreds of species and spatially long-range correlations among sites. Capturing this complex distribution requires longer training on the VAN, and thus the computational cost may grow with the number of species more than the nearly linear rate in the signaling cascade model. To further investigate the nonequilibrium phase transition needs to scan a set of values of the critical parameter \cite{tang2022solving}. Despite these challenges, since the computational cost of the present approach does not scale exponentially to the number of species, such studies become practically feasible by paralleling multiple GPUs. To this end, we have developed a user-friendly package extendable to stochastic reaction networks in more general situations. 

\clearpage
\section{Methods}

\subsection{Difference to the previous methods}

Compared with the Gillespie algorithm \cite{doob1942topics,gillespie1976general,weinan2021applied} of simulating trajectories, the present approach learns the joint probability distribution. 
Since the state space increases exponentially with the number of species, it is not feasible to extract the joint probability distribution from simulating trajectories for high-dimensional systems, for example, certain configurations may not be accessed unless an exponentially increasing number of trajectories are simulated (Supplementary Fig.~13). Instead, the joint probability distribution can be learned by the VAN, which offers direct sampling from the joint distribution to compute statistics, beyond acquiring only the marginal statistics from the Gillespie algorithm.  
Moreover, the trajectory space increases exponentially with the number of time points, making it harder to characterize the time evolution of the joint distribution by simulating trajectories and capturing rare trajectories. The present approach overcomes this issue by evolving the VAN to track the probability distribution over each time point. With the learnt time-evolving distribution covering the rare configurations, the dynamics under time-scale separations may be revealed.

The computational cost of the finite state projection \cite{munsky2006finite} and the sliding window method \cite{henzinger2009sliding} increases exponentially with the number of species, because the states need to be explicitly labeled to cast the CME into ordinary differential equations. The ACME method  \cite{cao2008optimal,cao2016accurate} also requires the buffers to be sufficiently large for accurate estimations, which may become computationally expensive when the upper count limit $N$ is large. Instead, the present approach captures the joint probability distribution of the whole state space with an adjustable $N$, and is efficient when both the upper count limit $N$ and the number of species $M$ are large, such as in the example of a signaling cascade. 
Given such a property, one can further conduct an estimation on choosing the proper $N$ to control the truncation error \cite{cao2016state}, and then $N$ can be adjusted accordingly for the neural network without losing the flexibility on learning the joint distribution. 
In addition, compared with the method based on the tensor network \cite{kazeev2014direct,ion2021tensor}, the advantages of the VAN mainly include its generality. The neural-network ansatz is more flexible representing complex probability distributions. For example, the VAN is useful in higher-dimensional lattice systems \cite{tang2022solving}, whereas the matrix product states are mainly for one-dimensional lattices \cite{causer2021finite}.

\subsection{Details on the VAN}

The details of implementing the VAN are as follows. For encoding the count of species, one may use the one-hot representation or binary representation. For the binary representation, each count of a species is encoded as the binary-digit representation. When using the VAN to sample configurations, the VAN maps binary digits back to $N$-output probabilities for each species. The required number of input variables for binary encoding increases logarithmically with the maximum count of species. 
Alternatively, to efficiently represent the count space by the least possible variables in the VAN, we directly used the count of each species in the range of $[0,N]$ as the input of the VAN. 
We next present two architectures to represent the probability in Eq.~\eqref{VAN}, the RNN and the transformer. 


\subsubsection{Recurrent neural network}

In the RNN, the conditional probability is calculated iteratively over species, with each site on the chain representing a species. Each site has two inputs: a visible variable with the count of species and a hidden state. The iteration process starts from an initial variable $n_{0}$ and hidden state $h_{0}$, which are chosen as zero. Then, a recurrent cell processes the information from the previous input variable $n_{i-1}$ and hidden state $h_{i-1}$, generates a new hidden state $h_{i}$, and passes the information of $h_{i}$ to the next cell. The recurrent cell also provides the conditional probability of this site  $\hat{P}^{\theta}(n_{i}|n_{1},\dots,n_{i-1})$ with the parameter $\theta$. The conditional probability is obtained from the hidden state, which is acted on a linear transform and a softmax operator $\hat{P}^{\theta}(n_{i}|n_{1},\dots,n_{i-1})=\text{Softmax}(Wh_{i}+b)$. The softmax operation ensures the normalized condition for the probability of all the possible counts of the species. The joint probability distribution is obtained by multiplying the iteratively generated conditional probabilities. 

Sampling from the probability distribution occurs in a similar manner. To sample configurations with discrete counts, we use the  multinomial distribution to generate integers based on the given probability. Given an initial variable and hidden state, the variable $n_{1}$ is sampled from the estimated conditional probability. The procedure is repeated to the last species, generating a configuration with a certain count of each species.

To learn the distribution with long-range correlations, we use a gated recurrent unit (GRU) \cite{cho2014properties} as the recurrent cell. It avoids the vanishing gradient problem for the vanilla RNN, and is more efficient than the cell of the long-short time memory. The GRU takes the $d_{h}$-dimensional hidden state $h_{i}$ as an input. It updates by the gates:
\begin{align}
z_{i}&=\sigma(W_{zn}n_{i-1}+W_{zh}h_{i-1}+b_{z}),\\
r_{i}&=\sigma(W_{rn}n_{i-1}+W_{rh}h_{i-1}+b_{r}),\\
\hat{h}_{i}&=\tanh(W_{hn}n_{i-1}+W_{hh}(r_{i}\odot h_{i-1})+b_{h}),\\
h_{i}&=(1-z_{i})\odot h_{i-1}+z_{i}\odot \hat{h}_{i},
\end{align}
where $W$s are the weight matrices,  $\sigma$ denotes the sigmoid activation function, $b$s are bias vectors, and $\odot$ is the Hadamard product. A reset gate $r_{i}$ and an update gate $z_{i}$ interpolate between the previous input hidden state and a candidate  hidden state, where the reset gate controls the extent of forgetting on the previous hidden state.  

To increase the representative capacity of the VAN, many hidden layers of RNN may be employed. We find that the one-layer RNN is sufficient to generate accurate probability distributions in examples. 

\subsubsection{Transformer}

The transformer \cite{NIPS2017_3f5ee243} employs the attention mechanism to learn the correlation between relevant configurations in a distribution. For the conventional transformer model, given a ``query'', it uses ``key'' to estimate a correlation to certain ``value'', and then generates the attention value. We implement  the transformer as a unit of the VAN, where the input is the count of each species and the output is the conditional probabilities in Eq.~\eqref{VAN}. 
Specifically, the transformer has an input layer, where a linear transform and the ReLU function act on the input configurations. Then, positional embedding is applied to the configurations, and the attention matrix for the configurations is calculated subsequently by using multihead attention \cite{NIPS2017_3f5ee243}. These are passed to a linear transform and then a softmax operator as in the RNN. The architecture ensures that the output conditional probabilities are normalized.

For the transformer, the hyperparameters include the number of features in the encoder and decoder inputs $d_{model}$,  the number of heads in the multihead attention models $n_{heads}$,  the number of encoder and decoder layers $n_{layers}$,  and the dimension of the feed-forward network model $d_{ff}$. The hyperparameter $n_{layers}$  plays a similar role as the number of depths in the RNN, and the other hyperparameters play the role of widths. The chosen values of hyperparameters are listed in Supplementary Table~\uppercase\expandafter{\romannumeral2}.

We have tested the transformer in the examples of the signaling cascade, where it shows a similar accuracy as the RNN. However, the transformer generally needs longer computational time than the RNN (Supplementary Table~\uppercase\expandafter{\romannumeral2}). Therefore, we have mainly used the RNN for other examples. The representative capacity of the transformer enables its application to more complex reaction networks. 

\subsection{The choice of upper count limit}

For a given problem, the upper count limit $N$ is typically fixed, because there is usually a lack of prior knowledge on the variation of the range over time. The choice of $N$ requires a special care: If $N$ is too small, the counting space is not fully covered; If $N$ is too large, training the VAN may require longer computational time to converge. When a priori knowledge of $N$  is unavailable, one can search for the smallest upper count limit such as by simulating trajectories from the Gillespie algorithm. Another more rigorous approach of finding the proper $N$  is to use the ACME method \cite{cao2016accurate,cao2016state}.  
This method based on the finite-buffer method  \cite{cao2008optimal} can reduce the state space from a $M$-dimensional hypercube to a $M$-simplex,  and provides an error bound on choosing $N$. It gives guidance on choosing an appropriate $N$ to truncate, ensuring that the truncation error is smaller than a threshold. After finding a proper $N$, the flexibility of neural networks in representing complex distributions allows $N$ to be adjusted accordingly in the present method. 

\subsection{Incorporating constraints to the VAN}

\subsubsection{Adding constraints on the count of each species}

When a priori knowledge on the count of certain chemical species is available, one can put separate constraints on the count  of each species, to contract the probability space and improve the accuracy of the VAN. For example, the DNA fragment typically has a count of zero or one inside a cell. Thus, one can constrain its count to $0, 1$, which reduces the state space of the probability distribution. 

We implement this constraint in the VAN, by setting the  probability of the exceeding  count in the output of the VAN to zero. That is, the count of the species out of the constraint is set to zero probability, such that only the allowable count is generated in the VAN. The constraint on the count can be specified for each species, as an input argument in our code package. It is exemplified in the toggle switch and the autoregulatory feedback loop.

\subsubsection{Keeping conservation on the total species count}

Many stochastic reaction networks in nature include a count conservation of species. Such conservation places a constraint on the total count of certain chemical  species. Therefore, the count conservation needs to be incorporated into the VAN. Here, we implement the total count conservation in the VAN similar as \cite{barrett2022autoregressive,PhysRevResearch.2.023358}, to model the chemical reactions with an intrinsic conserved quantity $N_{0}$. Such an implementation of count conservation is used in the example of an early life self-replicator.

To make the total count conserved, we adjust the output of each conditional probability in the VAN, to assign zero probabilities to the counts of species that do not satisfy the conservation. Specifically, the conditional probabilities $\hat{P}^{\theta}(n_{i}|n_{1},\dots,n_{i-1})$ for species $i$ are zero if 
\begin{align}
\label{Conservation}
0\leq n_{i}\leq N_{0}-\sum^{i-1}_{j=1}n_{j}    
\end{align} 
is not satisfied. This constraint ensures that the sampled configurations from the VAN must have the total count of species equal to $N_{0}$. We have used such strategies of adjusting the neural-network output for both sampling configurations and acquiring  probabilities from the VAN. The conservation of certain species can be realized in a similar manner.

We note that the conservation can also be maintained by performing the variable reduction. However, putting the constraint on the total conservation without the variable reduction is more straightforward in the VAN. For a high-dimensional system with many species under a constraint of $\sum^{M}_{i=1}n_{i}=N_{0}$, after performing the variable reduction, the remaining variables are still under the constraint of their counts and are dependent on the counts of other species.

\subsection{Training details}

Starting from a chosen initial distribution, we iteratively conduct the minimization by Eq.~\eqref{Loss}, generating a sequence of the VAN to continuously track the time evolution of the distribution. The first time step requires a long epoch ($\mathcal{O}(10^{3})$) to have a converged loss. Then, the learned VAN at the current time step is inherited to train the VAN after the next time-step evolution. After the first time step, the inherited VAN loss soon converges, because the change in the distribution is generally not dramatic. This inheritance significantly accelerates the training, taking $\mathcal{O}(10^{2})$ epochs at the time points after the first time step. 

The batch size is chosen as $1000$ for each epoch. To estimate the statistics from the distribution more accurately, after the loss converges, we save the batch from the last $10$ epochs. This collection of $1000\times10$ samples gives accurate estimations, and one can also save the VAN to generate more samples as desired (Supplementary Fig.~7). 
The Adam optimizer \cite{kingma2014adam} is chosen to perform the stochastic gradient descent, and the variance reduction is included to reduce the variance of the loss function \cite{PhysRevLett.122.080602}. We tested learning rates $10^{-5}$, $10^{-4}$, $10^{-3}$, and $10^{-2}$, which affect the accuracy of training. We find that $10^{-3}$  leads to relatively lower loss and better accuracy of training. Certain schedulers for the learning rate can be designed, to help alleviate general optimization issues such as trapping into local minima. 

The accuracy of the VAN depends on the number of parameters in the neural network. For the RNN, we have used $2, 4, 8, 16, 32, 64, 128$ as the number of hidden states, and found that the number larger than or equal to $16$ performs well for the one-layer RNN. A larger number of hidden states and hidden layers may lead to better accuracy, with a cost of longer computational time. We list the number of depths and widths of the RNN with the best performance under our attempt for each model (Table~\uppercase\expandafter{\romannumeral1}). The table contains the computational times in various systems with different numbers of chemical species and reactions, under the chosen time-step length. We also list the number of parameters for the transformer, which has equal performance but longer training time (Supplementary Table~\uppercase\expandafter{\romannumeral2}).

The VAN maps each count of species to the conditional probability of the species conditioned on the previous species. This procedure assigns a specific order to the species. Once the VAN can be trained to achieve zero KL-divergence, the joint distribution is faithfully learned regardless of the species order used in the conditional probabilities, because the parameters of the VAN are optimized to learn the joint probability distribution instead of each conditional probability. Specifically, we used different orders of the species for the conditional probabilities, such as shuffling the order of species in the example of the signaling cascade, and found that the learned probability distribution was not affected by the chosen order (Supplementary Fig.~9). If the chosen order respects the conditional dependence of species in the given reaction network, the training may be more efficient. A similar effect was analyzed in the context of supervised learning, where the optimal permutations on the order of variables can be searched by the spectral ordering \cite{acharya2022qubit}, to adapt the structure of learning models to the dataset.

To evaluate the accuracy of the learned distribution, we use two ways: examining the value of the converged loss function and the sample variance of the loss function (Supplementary Note~1); comparing the result  with other methods. First, the loss function of the KL-divergence quantifies the convergence and  accuracy of  learning the distribution. For example, the converged values of the KL-divergence and sample variance over time points are typically  below the order of $\mathcal{O}(10^{-4})$ in the toggle switch (Supplementary Fig.~1). It indicates that the variational training faithfully learns the distribution and does not suffer from a serious error accumulation over time. The  comparison with other methods are detailed in the examples.

 \subsection{The choice of time-step length}

The choice of the time-step length requires a care. It has the discretization error from the Suzuki-Trotter decomposition \cite{suzuki1976generalized} in Eq.~\eqref{Loss}. An improper choice of a too large time-step length may generate negative probabilities due to the escape rate in Eq.~\eqref{ME}. 
A sufficiently small time-step length is ideal for the accuracy, with a cost of longer computational time. Thus, the time-step length $\delta t$ needs to be adjusted to each given system, because the reaction rates have various values for each system. One needs to search for the proper $\delta t$ by trial and error. An empirically proper choice for each example is in Table~\ref{table1}. 

In addition to the fixed time-step length, one may use the option of adaptive step length in the code repository to save computational time. Specifically, by trial and error, we first find a sufficiently small step length as a baseline for a given system. 
Then, we increase the baseline value by $100$-fold. If it generates negative probabilities, the increase is reduced to $100/2$-fold, etc., until reaching the baseline of step length which does not give negative probabilities. This strategy allows us to explore the largest possible step length at each time point, to accelerate the simulation.

\textbf{Data availability:}
The authors declare that the data supporting this study are available within the paper.

\textbf{Code availability:}
A pytorch implementation of the present algorithm can be found at the code repository 
\newline 
https://github.com/jamestang23/NNCME \cite{Code}. 

\section*{Acknowledgments}
We thank Jing Liu for sharing the code of the transformer. We acknowledge Jie Liang, Mustafa Khammash, Ankit Gupta, Alexander Hoffmann, Ali Farhat, Farid Manuchehrfar and Online Club Nanothermodynamica for helpful discussions. 
This work is supported by Project 12105014 (Y.T.), 11747601 (P.Z.), 11975294 (P.Z.) of National Natural Science Foundation of China. 
P.Z. acknowledges the WIUCASICTP2022 grant. 
The HPC is supported by Dawning Information Industry Corporation Ltd and Interdisciplinary Intelligence SuperComputer Center of Beijing Normal University, Zhuhai.

\section*{Author contributions}
Y.T., P.Z. had the original idea for this work, Y.T. and J.W. performed the study, and all authors contributed to the preparation of the manuscript.

\section*{Competing interests} 

The authors declare no competing interests.

\section*{Additional information}


\textbf{Supplementary information} The online version
contains supplementary material available at [URL will be inserted by publisher].

\textbf{Correspondence and requests for materials} should be addressed to Ying Tang.

\textbf{Reprints and permission information}  is available online at  [URL will be inserted by publisher].


%


\begin{thebibliography}{54}%
\makeatletter
\providecommand \@ifxundefined [1]{%
 \@ifx{#1\undefined}
}%
\providecommand \@ifnum [1]{%
 \ifnum #1\expandafter \@firstoftwo
 \else \expandafter \@secondoftwo
 \fi
}%
\providecommand \@ifx [1]{%
 \ifx #1\expandafter \@firstoftwo
 \else \expandafter \@secondoftwo
 \fi
}%
\providecommand \natexlab [1]{#1}%
\providecommand \enquote  [1]{``#1''}%
\providecommand \bibnamefont  [1]{#1}%
\providecommand \bibfnamefont [1]{#1}%
\providecommand \citenamefont [1]{#1}%
\providecommand \href@noop [0]{\@secondoftwo}%
\providecommand \href [0]{\begingroup \@sanitize@url \@href}%
\providecommand \@href[1]{\@@startlink{#1}\@@href}%
\providecommand \@@href[1]{\endgroup#1\@@endlink}%
\providecommand \@sanitize@url [0]{\catcode `\\12\catcode `\$12\catcode
  `\&12\catcode `\#12\catcode `\^12\catcode `\_12\catcode `\%12\relax}%
\providecommand \@@startlink[1]{}%
\providecommand \@@endlink[0]{}%
\providecommand \url  [0]{\begingroup\@sanitize@url \@url }%
\providecommand \@url [1]{\endgroup\@href {#1}{\urlprefix }}%
\providecommand \urlprefix  [0]{URL }%
\providecommand \Eprint [0]{\href }%
\providecommand \doibase [0]{https://doi.org/}%
\providecommand \selectlanguage [0]{\@gobble}%
\providecommand \bibinfo  [0]{\@secondoftwo}%
\providecommand \bibfield  [0]{\@secondoftwo}%
\providecommand \translation [1]{[#1]}%
\providecommand \BibitemOpen [0]{}%
\providecommand \bibitemStop [0]{}%
\providecommand \bibitemNoStop [0]{.\EOS\space}%
\providecommand \EOS [0]{\spacefactor3000\relax}%
\providecommand \BibitemShut  [1]{\csname bibitem#1\endcsname}%
\let\auto@bib@innerbib\@empty
\bibitem [{\citenamefont {Weber}\ and\ \citenamefont
  {Frey}(2017)}]{weber2017master}%
  \BibitemOpen
  \bibfield  {author} {\bibinfo {author} {\bibfnamefont {M.~F.}\ \bibnamefont
  {Weber}}\ and\ \bibinfo {author} {\bibfnamefont {E.}~\bibnamefont {Frey}},\
  }\bibfield  {title} {\bibinfo {title} {Master equations and the theory of
  stochastic path integrals},\ }\href
  {https://iopscience.iop.org/article/10.1088/1361-6633/aa5ae2} {\bibfield
  {journal} {\bibinfo  {journal} {Rep. Prog. Phys.}\ }\textbf {\bibinfo
  {volume} {80}},\ \bibinfo {pages} {046601} (\bibinfo {year}
  {2017})}\BibitemShut {NoStop}%
\bibitem [{\citenamefont {Gillespie}(2007)}]{gillespie2007stochastic}%
  \BibitemOpen
  \bibfield  {author} {\bibinfo {author} {\bibfnamefont {D.~T.}\ \bibnamefont
  {Gillespie}},\ }\bibfield  {title} {\bibinfo {title} {Stochastic simulation
  of chemical kinetics},\ }\href
  {http://www.annualreviews.org/doi/abs/10.1146/annurev.physchem.58.032806.104637}
  {\bibfield  {journal} {\bibinfo  {journal} {Annu. Rev. Phys. Chem.}\ }\textbf
  {\bibinfo {volume} {58}},\ \bibinfo {pages} {35} (\bibinfo {year}
  {2007})}\BibitemShut {NoStop}%
\bibitem [{\citenamefont {Ge}\ \emph {et~al.}(2012)\citenamefont {Ge},
  \citenamefont {Qian},\ and\ \citenamefont {Qian}}]{ge2012stochastic}%
  \BibitemOpen
  \bibfield  {author} {\bibinfo {author} {\bibfnamefont {H.}~\bibnamefont
  {Ge}}, \bibinfo {author} {\bibfnamefont {M.}~\bibnamefont {Qian}},\ and\
  \bibinfo {author} {\bibfnamefont {H.}~\bibnamefont {Qian}},\ }\bibfield
  {title} {\bibinfo {title} {Stochastic theory of nonequilibrium steady states.
  part ii: Applications in chemical biophysics},\ }\href
  {http://www.sciencedirect.com/science/article/pii/S0370157311002419}
  {\bibfield  {journal} {\bibinfo  {journal} {Phys. Rep.}\ }\textbf {\bibinfo
  {volume} {510}},\ \bibinfo {pages} {87} (\bibinfo {year} {2012})}\BibitemShut
  {NoStop}%
\bibitem [{\citenamefont {Elowitz}\ \emph {et~al.}(2002)\citenamefont
  {Elowitz}, \citenamefont {Levine}, \citenamefont {Siggia},\ and\
  \citenamefont {Swain}}]{elowitz2002stochastic}%
  \BibitemOpen
  \bibfield  {author} {\bibinfo {author} {\bibfnamefont {M.~B.}\ \bibnamefont
  {Elowitz}}, \bibinfo {author} {\bibfnamefont {A.~J.}\ \bibnamefont {Levine}},
  \bibinfo {author} {\bibfnamefont {E.~D.}\ \bibnamefont {Siggia}},\ and\
  \bibinfo {author} {\bibfnamefont {P.~S.}\ \bibnamefont {Swain}},\ }\bibfield
  {title} {\bibinfo {title} {Stochastic gene expression in a single cell},\
  }\href {http://www.sciencemag.org/content/297/5584/1183.short} {\bibfield
  {journal} {\bibinfo  {journal} {Science}\ }\textbf {\bibinfo {volume}
  {297}},\ \bibinfo {pages} {1183} (\bibinfo {year} {2002})}\BibitemShut
  {NoStop}%
\bibitem [{\citenamefont {Blythe}\ and\ \citenamefont
  {McKane}(2007)}]{blythe2007stochastic}%
  \BibitemOpen
  \bibfield  {author} {\bibinfo {author} {\bibfnamefont {R.~A.}\ \bibnamefont
  {Blythe}}\ and\ \bibinfo {author} {\bibfnamefont {A.~J.}\ \bibnamefont
  {McKane}},\ }\bibfield  {title} {\bibinfo {title} {Stochastic models of
  evolution in genetics, ecology and linguistics},\ }\href
  {https://iopscience.iop.org/article/10.1088/1742-5468/2007/07/P07018/meta?casa_token=tYk6SCDMELwAAAAA:RfjocNxKlzz1cysneziriFNfehxljri1yCcDNQFJORXRgIt6p1olu5XOKqU4zXciDkvVPhzCMdftpNjH}
  {\bibfield  {journal} {\bibinfo  {journal} {J. Stat. Mech.}\ }\textbf
  {\bibinfo {volume} {2007}},\ \bibinfo {pages} {P07018} (\bibinfo {year}
  {2007})}\BibitemShut {NoStop}%
\bibitem [{\citenamefont {Jafarpour}\ \emph {et~al.}(2015)\citenamefont
  {Jafarpour}, \citenamefont {Biancalani},\ and\ \citenamefont
  {Goldenfeld}}]{PhysRevLett.115.158101}%
  \BibitemOpen
  \bibfield  {author} {\bibinfo {author} {\bibfnamefont {F.}~\bibnamefont
  {Jafarpour}}, \bibinfo {author} {\bibfnamefont {T.}~\bibnamefont
  {Biancalani}},\ and\ \bibinfo {author} {\bibfnamefont {N.}~\bibnamefont
  {Goldenfeld}},\ }\bibfield  {title} {\bibinfo {title} {Noise-induced
  mechanism for biological homochirality of early life self-replicators},\
  }\href {https://doi.org/10.1103/PhysRevLett.115.158101} {\bibfield  {journal}
  {\bibinfo  {journal} {Phys. Rev. Lett.}\ }\textbf {\bibinfo {volume} {115}},\
  \bibinfo {pages} {158101} (\bibinfo {year} {2015})}\BibitemShut {NoStop}%
\bibitem [{\citenamefont {Gardiner}(2004)}]{gardiner2004handbook}%
  \BibitemOpen
  \bibfield  {author} {\bibinfo {author} {\bibfnamefont {C.~W.}\ \bibnamefont
  {Gardiner}},\ }\href@noop {} {\emph {\bibinfo {title} {Handbook of Stochastic
  Methods}}},\ \bibinfo {edition} {3rd}\ ed.\ (\bibinfo  {publisher}
  {Springer-Verlag, Berlin},\ \bibinfo {year} {2004})\BibitemShut {NoStop}%
\bibitem [{\citenamefont {Frank}(1953)}]{frank1953spontaneous}%
  \BibitemOpen
  \bibfield  {author} {\bibinfo {author} {\bibfnamefont {F.~C.}\ \bibnamefont
  {Frank}},\ }\bibfield  {title} {\bibinfo {title} {On spontaneous asymmetric
  synthesis},\ }\href
  {https://www.sciencedirect.com/science/article/abs/pii/0006300253900821}
  {\bibfield  {journal} {\bibinfo  {journal} {Biochim. Biophys. Acta}\ }\textbf
  {\bibinfo {volume} {11}},\ \bibinfo {pages} {459} (\bibinfo {year}
  {1953})}\BibitemShut {NoStop}%
\bibitem [{\citenamefont {Bressloff}(2014)}]{bressloff2014stochastic}%
  \BibitemOpen
  \bibfield  {author} {\bibinfo {author} {\bibfnamefont {P.~C.}\ \bibnamefont
  {Bressloff}},\ }\href@noop {} {\emph {\bibinfo {title} {Stochastic Processes
  in Cell Biology}}},\ Vol.~\bibinfo {volume} {41}\ (\bibinfo  {publisher}
  {Springer, Berlin},\ \bibinfo {year} {2014})\BibitemShut {NoStop}%
\bibitem [{\citenamefont {Raj}\ and\ \citenamefont {van
  Oudenaarden}(2008)}]{raj2008nature}%
  \BibitemOpen
  \bibfield  {author} {\bibinfo {author} {\bibfnamefont {A.}~\bibnamefont
  {Raj}}\ and\ \bibinfo {author} {\bibfnamefont {A.}~\bibnamefont {van
  Oudenaarden}},\ }\bibfield  {title} {\bibinfo {title} {Nature, nurture, or
  chance: stochastic gene expression and its consequences},\ }\href
  {http://www.sciencedirect.com/science/article/pii/S0092867408012439}
  {\bibfield  {journal} {\bibinfo  {journal} {Cell}\ }\textbf {\bibinfo
  {volume} {135}},\ \bibinfo {pages} {216} (\bibinfo {year}
  {2008})}\BibitemShut {NoStop}%
\bibitem [{\citenamefont {van Kampen}(2007)}]{van2007stochastic}%
  \BibitemOpen
  \bibfield  {author} {\bibinfo {author} {\bibfnamefont {N.~G.}\ \bibnamefont
  {van Kampen}},\ }\href@noop {} {\emph {\bibinfo {title} {Stochastic Processes
  in Physics and Chemistry}}}\ (\bibinfo  {publisher} {Elsevier, New York},\
  \bibinfo {year} {2007})\BibitemShut {NoStop}%
\bibitem [{\citenamefont {Doob}(1942)}]{doob1942topics}%
  \BibitemOpen
  \bibfield  {author} {\bibinfo {author} {\bibfnamefont {J.~L.}\ \bibnamefont
  {Doob}},\ }\bibfield  {title} {\bibinfo {title} {Topics in the theory of
  markoff chains},\ }\href {https://www.jstor.org/stable/1990152} {\bibfield
  {journal} {\bibinfo  {journal} {Trans. Am. Math. Soc.}\ }\textbf {\bibinfo
  {volume} {52}},\ \bibinfo {pages} {37} (\bibinfo {year} {1942})}\BibitemShut
  {NoStop}%
\bibitem [{\citenamefont {Gillespie}(1976)}]{gillespie1976general}%
  \BibitemOpen
  \bibfield  {author} {\bibinfo {author} {\bibfnamefont {D.~T.}\ \bibnamefont
  {Gillespie}},\ }\bibfield  {title} {\bibinfo {title} {A general method for
  numerically simulating the stochastic time evolution of coupled chemical
  reactions},\ }\href
  {http://www.sciencedirect.com/science/article/pii/0021999176900413}
  {\bibfield  {journal} {\bibinfo  {journal} {J. Comput. Phys.}\ }\textbf
  {\bibinfo {volume} {22}},\ \bibinfo {pages} {403} (\bibinfo {year}
  {1976})}\BibitemShut {NoStop}%
\bibitem [{\citenamefont {Weinan}\ \emph {et~al.}(2021)\citenamefont {Weinan},
  \citenamefont {Li},\ and\ \citenamefont
  {Vanden-Eijnden}}]{weinan2021applied}%
  \BibitemOpen
  \bibfield  {author} {\bibinfo {author} {\bibfnamefont {E.}~\bibnamefont
  {Weinan}}, \bibinfo {author} {\bibfnamefont {T.}~\bibnamefont {Li}},\ and\
  \bibinfo {author} {\bibfnamefont {E.}~\bibnamefont {Vanden-Eijnden}},\
  }\href@noop {} {\emph {\bibinfo {title} {Applied Stochastic Analysis}}},\
  Vol.\ \bibinfo {volume} {199}\ (\bibinfo  {publisher} {American Mathematical
  Society},\ \bibinfo {year} {2021})\BibitemShut {NoStop}%
\bibitem [{\citenamefont {Terebus}\ \emph {et~al.}(2019)\citenamefont
  {Terebus}, \citenamefont {Liu},\ and\ \citenamefont
  {Liang}}]{terebus2019discrete}%
  \BibitemOpen
  \bibfield  {author} {\bibinfo {author} {\bibfnamefont {A.}~\bibnamefont
  {Terebus}}, \bibinfo {author} {\bibfnamefont {C.}~\bibnamefont {Liu}},\ and\
  \bibinfo {author} {\bibfnamefont {J.}~\bibnamefont {Liang}},\ }\bibfield
  {title} {\bibinfo {title} {Discrete and continuous models of probability flux
  of switching dynamics: Uncovering stochastic oscillations in a toggle-switch
  system},\ }\href {https://doi.org/10.1063/1.5124823} {\bibfield  {journal}
  {\bibinfo  {journal} {J. Chem. Phys.}\ }\textbf {\bibinfo {volume} {151}},\
  \bibinfo {pages} {185104} (\bibinfo {year} {2019})}\BibitemShut {NoStop}%
\bibitem [{\citenamefont {Terebus}\ \emph {et~al.}(2021)\citenamefont
  {Terebus}, \citenamefont {Manuchehrfar}, \citenamefont {Cao},\ and\
  \citenamefont {Liang}}]{terebus2021exact}%
  \BibitemOpen
  \bibfield  {author} {\bibinfo {author} {\bibfnamefont {A.}~\bibnamefont
  {Terebus}}, \bibinfo {author} {\bibfnamefont {F.}~\bibnamefont
  {Manuchehrfar}}, \bibinfo {author} {\bibfnamefont {Y.}~\bibnamefont {Cao}},\
  and\ \bibinfo {author} {\bibfnamefont {J.}~\bibnamefont {Liang}},\ }\bibfield
   {title} {\bibinfo {title} {Exact probability landscapes of stochastic
  phenotype switching in feed-forward loops: Phase diagrams of multimodality},\
  }\href {https://www.frontiersin.org/articles/10.3389/fgene.2021.645640/full}
  {\bibfield  {journal} {\bibinfo  {journal} {Front. Genet.}\ }\textbf
  {\bibinfo {volume} {12}},\ \bibinfo {pages} {645640} (\bibinfo {year}
  {2021})}\BibitemShut {NoStop}%
\bibitem [{\citenamefont {Gillespie}(2000)}]{gillespie2000chemical}%
  \BibitemOpen
  \bibfield  {author} {\bibinfo {author} {\bibfnamefont {D.~T.}\ \bibnamefont
  {Gillespie}},\ }\bibfield  {title} {\bibinfo {title} {The chemical langevin
  equation},\ }\href
  {http://scitation.aip.org/content/aip/journal/jcp/113/1/10.1063/1.481811}
  {\bibfield  {journal} {\bibinfo  {journal} {J. Chem. Phys.}\ }\textbf
  {\bibinfo {volume} {113}},\ \bibinfo {pages} {297} (\bibinfo {year}
  {2000})}\BibitemShut {NoStop}%
\bibitem [{\citenamefont {Munsky}\ and\ \citenamefont
  {Khammash}(2006)}]{munsky2006finite}%
  \BibitemOpen
  \bibfield  {author} {\bibinfo {author} {\bibfnamefont {B.}~\bibnamefont
  {Munsky}}\ and\ \bibinfo {author} {\bibfnamefont {M.}~\bibnamefont
  {Khammash}},\ }\bibfield  {title} {\bibinfo {title} {The finite state
  projection algorithm for the solution of the chemical master equation},\
  }\href {https://aip.scitation.org/doi/full/10.1063/1.2145882} {\bibfield
  {journal} {\bibinfo  {journal} {J. Chem. Phys.}\ }\textbf {\bibinfo {volume}
  {124}},\ \bibinfo {pages} {044104} (\bibinfo {year} {2006})}\BibitemShut
  {NoStop}%
\bibitem [{\citenamefont {Henzinger}\ \emph {et~al.}(2009)\citenamefont
  {Henzinger}, \citenamefont {Mateescu},\ and\ \citenamefont
  {Wolf}}]{henzinger2009sliding}%
  \BibitemOpen
  \bibfield  {author} {\bibinfo {author} {\bibfnamefont {T.~A.}\ \bibnamefont
  {Henzinger}}, \bibinfo {author} {\bibfnamefont {M.}~\bibnamefont
  {Mateescu}},\ and\ \bibinfo {author} {\bibfnamefont {V.}~\bibnamefont
  {Wolf}},\ }\bibfield  {title} {\bibinfo {title} {Sliding window abstraction
  for infinite markov chains},\ }in\ \href
  {https://link.springer.com/chapter/10.1007/978-3-642-02658-4_27} {\emph
  {\bibinfo {booktitle} {Computer Aided Verification}}}\ (\bibinfo
  {organization} {Springer},\ \bibinfo {year} {2009})\ pp.\ \bibinfo {pages}
  {337--352}\BibitemShut {NoStop}%
\bibitem [{\citenamefont {Cao}\ \emph {et~al.}(2016{\natexlab{a}})\citenamefont
  {Cao}, \citenamefont {Terebus},\ and\ \citenamefont
  {Liang}}]{cao2016accurate}%
  \BibitemOpen
  \bibfield  {author} {\bibinfo {author} {\bibfnamefont {Y.}~\bibnamefont
  {Cao}}, \bibinfo {author} {\bibfnamefont {A.}~\bibnamefont {Terebus}},\ and\
  \bibinfo {author} {\bibfnamefont {J.}~\bibnamefont {Liang}},\ }\bibfield
  {title} {\bibinfo {title} {Accurate chemical master equation solution using
  multi-finite buffers},\ }\href
  {https://epubs.siam.org/doi/abs/10.1137/15M1034180?casa_token=RxIozl4VSf0AAAAA:JInUD-ee6_TGqxwYXrp-4a6yQP22ceDsIMLNWUpSbDaaV38GOSBG36UDebCeVIx5EmxjKKLsRHc}
  {\bibfield  {journal} {\bibinfo  {journal} {Multiscale Model. Simul.}\
  }\textbf {\bibinfo {volume} {14}},\ \bibinfo {pages} {923} (\bibinfo {year}
  {2016}{\natexlab{a}})}\BibitemShut {NoStop}%
\bibitem [{\citenamefont {Cao}\ \emph {et~al.}(2016{\natexlab{b}})\citenamefont
  {Cao}, \citenamefont {Terebus},\ and\ \citenamefont {Liang}}]{cao2016state}%
  \BibitemOpen
  \bibfield  {author} {\bibinfo {author} {\bibfnamefont {Y.}~\bibnamefont
  {Cao}}, \bibinfo {author} {\bibfnamefont {A.}~\bibnamefont {Terebus}},\ and\
  \bibinfo {author} {\bibfnamefont {J.}~\bibnamefont {Liang}},\ }\bibfield
  {title} {\bibinfo {title} {State space truncation with quantified errors for
  accurate solutions to discrete chemical master equation},\ }\href
  {https://link.springer.com/article/10.1007/s11538-016-0149-1} {\bibfield
  {journal} {\bibinfo  {journal} {Bull. Math. Biol.}\ }\textbf {\bibinfo
  {volume} {78}},\ \bibinfo {pages} {617} (\bibinfo {year}
  {2016}{\natexlab{b}})}\BibitemShut {NoStop}%
\bibitem [{\citenamefont {MacNamara}\ \emph {et~al.}(2008)\citenamefont
  {MacNamara}, \citenamefont {Burrage},\ and\ \citenamefont
  {Sidje}}]{doi:10.1137/060678154}%
  \BibitemOpen
  \bibfield  {author} {\bibinfo {author} {\bibfnamefont {S.}~\bibnamefont
  {MacNamara}}, \bibinfo {author} {\bibfnamefont {K.}~\bibnamefont {Burrage}},\
  and\ \bibinfo {author} {\bibfnamefont {R.~B.}\ \bibnamefont {Sidje}},\
  }\bibfield  {title} {\bibinfo {title} {Multiscale modeling of chemical
  kinetics via the master equation},\ }\href
  {https://doi.org/10.1137/060678154} {\bibfield  {journal} {\bibinfo
  {journal} {Multiscale Model. Simul.}\ }\textbf {\bibinfo {volume} {6}},\
  \bibinfo {pages} {1146} (\bibinfo {year} {2008})}\BibitemShut {NoStop}%
\bibitem [{\citenamefont {Kazeev}\ \emph {et~al.}(2014)\citenamefont {Kazeev},
  \citenamefont {Khammash}, \citenamefont {Nip},\ and\ \citenamefont
  {Schwab}}]{kazeev2014direct}%
  \BibitemOpen
  \bibfield  {author} {\bibinfo {author} {\bibfnamefont {V.}~\bibnamefont
  {Kazeev}}, \bibinfo {author} {\bibfnamefont {M.}~\bibnamefont {Khammash}},
  \bibinfo {author} {\bibfnamefont {M.}~\bibnamefont {Nip}},\ and\ \bibinfo
  {author} {\bibfnamefont {C.}~\bibnamefont {Schwab}},\ }\bibfield  {title}
  {\bibinfo {title} {Direct solution of the chemical master equation using
  quantized tensor trains},\ }\href
  {https://journals.plos.org/ploscompbiol/article?id=10.1371/journal.pcbi.1003359}
  {\bibfield  {journal} {\bibinfo  {journal} {PLoS Comput. Biol.}\ }\textbf
  {\bibinfo {volume} {10}},\ \bibinfo {pages} {e1003359} (\bibinfo {year}
  {2014})}\BibitemShut {NoStop}%
\bibitem [{\citenamefont {Ion}\ \emph {et~al.}(2021)\citenamefont {Ion},
  \citenamefont {Wildner}, \citenamefont {Loukrezis}, \citenamefont {Koeppl},\
  and\ \citenamefont {De~Gersem}}]{ion2021tensor}%
  \BibitemOpen
  \bibfield  {author} {\bibinfo {author} {\bibfnamefont {I.~G.}\ \bibnamefont
  {Ion}}, \bibinfo {author} {\bibfnamefont {C.}~\bibnamefont {Wildner}},
  \bibinfo {author} {\bibfnamefont {D.}~\bibnamefont {Loukrezis}}, \bibinfo
  {author} {\bibfnamefont {H.}~\bibnamefont {Koeppl}},\ and\ \bibinfo {author}
  {\bibfnamefont {H.}~\bibnamefont {De~Gersem}},\ }\bibfield  {title} {\bibinfo
  {title} {Tensor-train approximation of the chemical master equation and its
  application for parameter inference},\ }\href
  {https://aip.scitation.org/doi/full/10.1063/5.0045521} {\bibfield  {journal}
  {\bibinfo  {journal} {J. Chem. Phys.}\ }\textbf {\bibinfo {volume} {155}},\
  \bibinfo {pages} {034102} (\bibinfo {year} {2021})}\BibitemShut {NoStop}%
\bibitem [{\citenamefont {Gupta}\ \emph {et~al.}(2021)\citenamefont {Gupta},
  \citenamefont {Schwab},\ and\ \citenamefont {Khammash}}]{gupta2021deepcme}%
  \BibitemOpen
  \bibfield  {author} {\bibinfo {author} {\bibfnamefont {A.}~\bibnamefont
  {Gupta}}, \bibinfo {author} {\bibfnamefont {C.}~\bibnamefont {Schwab}},\ and\
  \bibinfo {author} {\bibfnamefont {M.}~\bibnamefont {Khammash}},\ }\bibfield
  {title} {\bibinfo {title} {Deepcme: A deep learning framework for computing
  solution statistics of the chemical master equation},\ }\href
  {https://journals.plos.org/ploscompbiol/article?id=10.1371/journal.pcbi.1009623}
  {\bibfield  {journal} {\bibinfo  {journal} {PLoS Comput. Biol.}\ }\textbf
  {\bibinfo {volume} {17}},\ \bibinfo {pages} {e1009623} (\bibinfo {year}
  {2021})}\BibitemShut {NoStop}%
\bibitem [{\citenamefont {Mehta}\ \emph {et~al.}(2019)\citenamefont {Mehta},
  \citenamefont {Bukov}, \citenamefont {Wang}, \citenamefont {Day},
  \citenamefont {Richardson}, \citenamefont {Fisher},\ and\ \citenamefont
  {Schwab}}]{mehta2019high}%
  \BibitemOpen
  \bibfield  {author} {\bibinfo {author} {\bibfnamefont {P.}~\bibnamefont
  {Mehta}}, \bibinfo {author} {\bibfnamefont {M.}~\bibnamefont {Bukov}},
  \bibinfo {author} {\bibfnamefont {C.-H.}\ \bibnamefont {Wang}}, \bibinfo
  {author} {\bibfnamefont {A.~G.}\ \bibnamefont {Day}}, \bibinfo {author}
  {\bibfnamefont {C.}~\bibnamefont {Richardson}}, \bibinfo {author}
  {\bibfnamefont {C.~K.}\ \bibnamefont {Fisher}},\ and\ \bibinfo {author}
  {\bibfnamefont {D.~J.}\ \bibnamefont {Schwab}},\ }\bibfield  {title}
  {\bibinfo {title} {A high-bias, low-variance introduction to machine learning
  for physicists},\ }\href
  {https://www.sciencedirect.com/science/article/pii/S0370157319300766}
  {\bibfield  {journal} {\bibinfo  {journal} {Phys. Rep.}\ } (\bibinfo {year}
  {2019})}\BibitemShut {NoStop}%
\bibitem [{\citenamefont {Carleo}\ \emph {et~al.}(2019)\citenamefont {Carleo},
  \citenamefont {Cirac}, \citenamefont {Cranmer}, \citenamefont {Daudet},
  \citenamefont {Schuld}, \citenamefont {Tishby}, \citenamefont
  {Vogt-Maranto},\ and\ \citenamefont {Zdeborov\'a}}]{RevModPhys.91.045002}%
  \BibitemOpen
  \bibfield  {author} {\bibinfo {author} {\bibfnamefont {G.}~\bibnamefont
  {Carleo}}, \bibinfo {author} {\bibfnamefont {I.}~\bibnamefont {Cirac}},
  \bibinfo {author} {\bibfnamefont {K.}~\bibnamefont {Cranmer}}, \bibinfo
  {author} {\bibfnamefont {L.}~\bibnamefont {Daudet}}, \bibinfo {author}
  {\bibfnamefont {M.}~\bibnamefont {Schuld}}, \bibinfo {author} {\bibfnamefont
  {N.}~\bibnamefont {Tishby}}, \bibinfo {author} {\bibfnamefont
  {L.}~\bibnamefont {Vogt-Maranto}},\ and\ \bibinfo {author} {\bibfnamefont
  {L.}~\bibnamefont {Zdeborov\'a}},\ }\bibfield  {title} {\bibinfo {title}
  {Machine learning and the physical sciences},\ }\href
  {https://doi.org/10.1103/RevModPhys.91.045002} {\bibfield  {journal}
  {\bibinfo  {journal} {Rev. Mod. Phys.}\ }\textbf {\bibinfo {volume} {91}},\
  \bibinfo {pages} {045002} (\bibinfo {year} {2019})}\BibitemShut {NoStop}%
\bibitem [{\citenamefont {Tang}\ and\ \citenamefont
  {Hoffmann}(2022)}]{Tang_2022}%
  \BibitemOpen
  \bibfield  {author} {\bibinfo {author} {\bibfnamefont {Y.}~\bibnamefont
  {Tang}}\ and\ \bibinfo {author} {\bibfnamefont {A.}~\bibnamefont
  {Hoffmann}},\ }\bibfield  {title} {\bibinfo {title} {Quantifying information
  of intracellular signaling: progress with machine learning},\ }\href
  {https://doi.org/10.1088/1361-6633/ac7a4a} {\bibfield  {journal} {\bibinfo
  {journal} {Rep. Prog. Phys.}\ }\textbf {\bibinfo {volume} {85}},\ \bibinfo
  {pages} {086602} (\bibinfo {year} {2022})}\BibitemShut {NoStop}%
\bibitem [{\citenamefont {Wu}\ \emph {et~al.}(2019)\citenamefont {Wu},
  \citenamefont {Wang},\ and\ \citenamefont {Zhang}}]{PhysRevLett.122.080602}%
  \BibitemOpen
  \bibfield  {author} {\bibinfo {author} {\bibfnamefont {D.}~\bibnamefont
  {Wu}}, \bibinfo {author} {\bibfnamefont {L.}~\bibnamefont {Wang}},\ and\
  \bibinfo {author} {\bibfnamefont {P.}~\bibnamefont {Zhang}},\ }\bibfield
  {title} {\bibinfo {title} {Solving statistical mechanics using variational
  autoregressive networks},\ }\href
  {https://doi.org/10.1103/PhysRevLett.122.080602} {\bibfield  {journal}
  {\bibinfo  {journal} {Phys. Rev. Lett.}\ }\textbf {\bibinfo {volume} {122}},\
  \bibinfo {pages} {080602} (\bibinfo {year} {2019})}\BibitemShut {NoStop}%
\bibitem [{\citenamefont {Hibat-Allah}\ \emph {et~al.}(2020)\citenamefont
  {Hibat-Allah}, \citenamefont {Ganahl}, \citenamefont {Hayward}, \citenamefont
  {Melko},\ and\ \citenamefont {Carrasquilla}}]{PhysRevResearch.2.023358}%
  \BibitemOpen
  \bibfield  {author} {\bibinfo {author} {\bibfnamefont {M.}~\bibnamefont
  {Hibat-Allah}}, \bibinfo {author} {\bibfnamefont {M.}~\bibnamefont {Ganahl}},
  \bibinfo {author} {\bibfnamefont {L.~E.}\ \bibnamefont {Hayward}}, \bibinfo
  {author} {\bibfnamefont {R.~G.}\ \bibnamefont {Melko}},\ and\ \bibinfo
  {author} {\bibfnamefont {J.}~\bibnamefont {Carrasquilla}},\ }\bibfield
  {title} {\bibinfo {title} {Recurrent neural network wave functions},\ }\href
  {https://doi.org/10.1103/PhysRevResearch.2.023358} {\bibfield  {journal}
  {\bibinfo  {journal} {Phys. Rev. Research}\ }\textbf {\bibinfo {volume}
  {2}},\ \bibinfo {pages} {023358} (\bibinfo {year} {2020})}\BibitemShut
  {NoStop}%
\bibitem [{\citenamefont {Sharir}\ \emph {et~al.}(2020)\citenamefont {Sharir},
  \citenamefont {Levine}, \citenamefont {Wies}, \citenamefont {Carleo},\ and\
  \citenamefont {Shashua}}]{PhysRevLett.124.020503}%
  \BibitemOpen
  \bibfield  {author} {\bibinfo {author} {\bibfnamefont {O.}~\bibnamefont
  {Sharir}}, \bibinfo {author} {\bibfnamefont {Y.}~\bibnamefont {Levine}},
  \bibinfo {author} {\bibfnamefont {N.}~\bibnamefont {Wies}}, \bibinfo {author}
  {\bibfnamefont {G.}~\bibnamefont {Carleo}},\ and\ \bibinfo {author}
  {\bibfnamefont {A.}~\bibnamefont {Shashua}},\ }\bibfield  {title} {\bibinfo
  {title} {Deep autoregressive models for the efficient variational simulation
  of many-body quantum systems},\ }\href
  {https://doi.org/10.1103/PhysRevLett.124.020503} {\bibfield  {journal}
  {\bibinfo  {journal} {Phys. Rev. Lett.}\ }\textbf {\bibinfo {volume} {124}},\
  \bibinfo {pages} {020503} (\bibinfo {year} {2020})}\BibitemShut {NoStop}%
\bibitem [{\citenamefont {Barrett}\ \emph {et~al.}(2022)\citenamefont
  {Barrett}, \citenamefont {Malyshev},\ and\ \citenamefont
  {Lvovsky}}]{barrett2022autoregressive}%
  \BibitemOpen
  \bibfield  {author} {\bibinfo {author} {\bibfnamefont {T.~D.}\ \bibnamefont
  {Barrett}}, \bibinfo {author} {\bibfnamefont {A.}~\bibnamefont {Malyshev}},\
  and\ \bibinfo {author} {\bibfnamefont {A.}~\bibnamefont {Lvovsky}},\
  }\bibfield  {title} {\bibinfo {title} {Autoregressive neural-network
  wavefunctions for ab initio quantum chemistry},\ }\href
  {https://www.nature.com/articles/s42256-022-00461-z} {\bibfield  {journal}
  {\bibinfo  {journal} {Nat. Mach. Intell.}\ }\textbf {\bibinfo {volume} {4}},\
  \bibinfo {pages} {351} (\bibinfo {year} {2022})}\BibitemShut {NoStop}%
\bibitem [{\citenamefont {Luo}\ \emph {et~al.}(2022)\citenamefont {Luo},
  \citenamefont {Chen}, \citenamefont {Carrasquilla},\ and\ \citenamefont
  {Clark}}]{PhysRevLett.128.090501}%
  \BibitemOpen
  \bibfield  {author} {\bibinfo {author} {\bibfnamefont {D.}~\bibnamefont
  {Luo}}, \bibinfo {author} {\bibfnamefont {Z.}~\bibnamefont {Chen}}, \bibinfo
  {author} {\bibfnamefont {J.}~\bibnamefont {Carrasquilla}},\ and\ \bibinfo
  {author} {\bibfnamefont {B.~K.}\ \bibnamefont {Clark}},\ }\bibfield  {title}
  {\bibinfo {title} {Autoregressive neural network for simulating open quantum
  systems via a probabilistic formulation},\ }\href
  {https://doi.org/10.1103/PhysRevLett.128.090501} {\bibfield  {journal}
  {\bibinfo  {journal} {Phys. Rev. Lett.}\ }\textbf {\bibinfo {volume} {128}},\
  \bibinfo {pages} {090501} (\bibinfo {year} {2022})}\BibitemShut {NoStop}%
\bibitem [{\citenamefont {Carrasquilla}\ \emph {et~al.}(2021)\citenamefont
  {Carrasquilla}, \citenamefont {Luo}, \citenamefont {P\'erez}, \citenamefont
  {Milsted}, \citenamefont {Clark}, \citenamefont {Volkovs},\ and\
  \citenamefont {Aolita}}]{PhysRevA.104.032610}%
  \BibitemOpen
  \bibfield  {author} {\bibinfo {author} {\bibfnamefont {J.}~\bibnamefont
  {Carrasquilla}}, \bibinfo {author} {\bibfnamefont {D.}~\bibnamefont {Luo}},
  \bibinfo {author} {\bibfnamefont {F.}~\bibnamefont {P\'erez}}, \bibinfo
  {author} {\bibfnamefont {A.}~\bibnamefont {Milsted}}, \bibinfo {author}
  {\bibfnamefont {B.~K.}\ \bibnamefont {Clark}}, \bibinfo {author}
  {\bibfnamefont {M.}~\bibnamefont {Volkovs}},\ and\ \bibinfo {author}
  {\bibfnamefont {L.}~\bibnamefont {Aolita}},\ }\bibfield  {title} {\bibinfo
  {title} {Probabilistic simulation of quantum circuits using a deep-learning
  architecture},\ }\href {https://doi.org/10.1103/PhysRevA.104.032610}
  {\bibfield  {journal} {\bibinfo  {journal} {Phys. Rev. A}\ }\textbf {\bibinfo
  {volume} {104}},\ \bibinfo {pages} {032610} (\bibinfo {year}
  {2021})}\BibitemShut {NoStop}%
\bibitem [{\citenamefont {Shin}\ \emph {et~al.}(2021)\citenamefont {Shin},
  \citenamefont {Riesselman}, \citenamefont {Kollasch}, \citenamefont
  {McMahon}, \citenamefont {Simon}, \citenamefont {Sander}, \citenamefont
  {Manglik}, \citenamefont {Kruse},\ and\ \citenamefont
  {Marks}}]{shin2021protein}%
  \BibitemOpen
  \bibfield  {author} {\bibinfo {author} {\bibfnamefont {J.-E.}\ \bibnamefont
  {Shin}}, \bibinfo {author} {\bibfnamefont {A.~J.}\ \bibnamefont
  {Riesselman}}, \bibinfo {author} {\bibfnamefont {A.~W.}\ \bibnamefont
  {Kollasch}}, \bibinfo {author} {\bibfnamefont {C.}~\bibnamefont {McMahon}},
  \bibinfo {author} {\bibfnamefont {E.}~\bibnamefont {Simon}}, \bibinfo
  {author} {\bibfnamefont {C.}~\bibnamefont {Sander}}, \bibinfo {author}
  {\bibfnamefont {A.}~\bibnamefont {Manglik}}, \bibinfo {author} {\bibfnamefont
  {A.~C.}\ \bibnamefont {Kruse}},\ and\ \bibinfo {author} {\bibfnamefont
  {D.~S.}\ \bibnamefont {Marks}},\ }\bibfield  {title} {\bibinfo {title}
  {Protein design and variant prediction using autoregressive generative
  models},\ }\href {https://www.nature.com/articles/s41467-021-22732-w}
  {\bibfield  {journal} {\bibinfo  {journal} {Nat. Commun.}\ }\textbf {\bibinfo
  {volume} {12}},\ \bibinfo {pages} {2403} (\bibinfo {year}
  {2021})}\BibitemShut {NoStop}%
\bibitem [{\citenamefont {Vaswani}\ \emph {et~al.}(2017)\citenamefont
  {Vaswani}, \citenamefont {Shazeer}, \citenamefont {Parmar}, \citenamefont
  {Uszkoreit}, \citenamefont {Jones}, \citenamefont {Gomez}, \citenamefont
  {Kaiser},\ and\ \citenamefont {Polosukhin}}]{NIPS2017_3f5ee243}%
  \BibitemOpen
  \bibfield  {author} {\bibinfo {author} {\bibfnamefont {A.}~\bibnamefont
  {Vaswani}}, \bibinfo {author} {\bibfnamefont {N.}~\bibnamefont {Shazeer}},
  \bibinfo {author} {\bibfnamefont {N.}~\bibnamefont {Parmar}}, \bibinfo
  {author} {\bibfnamefont {J.}~\bibnamefont {Uszkoreit}}, \bibinfo {author}
  {\bibfnamefont {L.}~\bibnamefont {Jones}}, \bibinfo {author} {\bibfnamefont
  {A.~N.}\ \bibnamefont {Gomez}}, \bibinfo {author} {\bibfnamefont {L.~u.}\
  \bibnamefont {Kaiser}},\ and\ \bibinfo {author} {\bibfnamefont
  {I.}~\bibnamefont {Polosukhin}},\ }\bibfield  {title} {\bibinfo {title}
  {Attention is all you need},\ }in\ \href
  {https://proceedings.neurips.cc/paper/2017/file/3f5ee243547dee91fbd053c1c4a845aa-Paper.pdf}
  {\emph {\bibinfo {booktitle} {Advances in Neural Information Processing
  Systems}}},\ Vol.~\bibinfo {volume} {30},\ \bibinfo {editor} {edited by\
  \bibinfo {editor} {\bibfnamefont {I.}~\bibnamefont {Guyon}}, \bibinfo
  {editor} {\bibfnamefont {U.~V.}\ \bibnamefont {Luxburg}}, \bibinfo {editor}
  {\bibfnamefont {S.}~\bibnamefont {Bengio}}, \bibinfo {editor} {\bibfnamefont
  {H.}~\bibnamefont {Wallach}}, \bibinfo {editor} {\bibfnamefont
  {R.}~\bibnamefont {Fergus}}, \bibinfo {editor} {\bibfnamefont
  {S.}~\bibnamefont {Vishwanathan}},\ and\ \bibinfo {editor} {\bibfnamefont
  {R.}~\bibnamefont {Garnett}}}\ (\bibinfo  {publisher} {Curran Associates,
  Inc.},\ \bibinfo {year} {2017})\BibitemShut {NoStop}%
\bibitem [{\citenamefont {Jiang}\ \emph {et~al.}(2021)\citenamefont {Jiang},
  \citenamefont {Fu}, \citenamefont {Yan}, \citenamefont {Li}, \citenamefont
  {Du}, \citenamefont {Cao}, \citenamefont {Qian},\ and\ \citenamefont
  {Grima}}]{jiang2021neural}%
  \BibitemOpen
  \bibfield  {author} {\bibinfo {author} {\bibfnamefont {Q.}~\bibnamefont
  {Jiang}}, \bibinfo {author} {\bibfnamefont {X.}~\bibnamefont {Fu}}, \bibinfo
  {author} {\bibfnamefont {S.}~\bibnamefont {Yan}}, \bibinfo {author}
  {\bibfnamefont {R.}~\bibnamefont {Li}}, \bibinfo {author} {\bibfnamefont
  {W.}~\bibnamefont {Du}}, \bibinfo {author} {\bibfnamefont {Z.}~\bibnamefont
  {Cao}}, \bibinfo {author} {\bibfnamefont {F.}~\bibnamefont {Qian}},\ and\
  \bibinfo {author} {\bibfnamefont {R.}~\bibnamefont {Grima}},\ }\bibfield
  {title} {\bibinfo {title} {Neural network aided approximation and parameter
  inference of non-markovian models of gene expression},\ }\href
  {https://www.nature.com/articles/s41467-021-22919-1} {\bibfield  {journal}
  {\bibinfo  {journal} {Nat. Commun.}\ }\textbf {\bibinfo {volume} {12}},\
  \bibinfo {pages} {1} (\bibinfo {year} {2021})}\BibitemShut {NoStop}%
\bibitem [{\citenamefont {Sukys}\ \emph {et~al.}(2022)\citenamefont {Sukys},
  \citenamefont {{\"O}cal},\ and\ \citenamefont
  {Grima}}]{sukys2022approximating}%
  \BibitemOpen
  \bibfield  {author} {\bibinfo {author} {\bibfnamefont {A.}~\bibnamefont
  {Sukys}}, \bibinfo {author} {\bibfnamefont {K.}~\bibnamefont {{\"O}cal}},\
  and\ \bibinfo {author} {\bibfnamefont {R.}~\bibnamefont {Grima}},\ }\bibfield
   {title} {\bibinfo {title} {Approximating solutions of the chemical master
  equation using neural networks},\ }\href
  {https://doi.org/https://doi.org/10.1016/j.isci.2022.105010} {\bibfield
  {journal} {\bibinfo  {journal} {iScience}\ }\textbf {\bibinfo {volume}
  {25}},\ \bibinfo {pages} {105010} (\bibinfo {year} {2022})}\BibitemShut
  {NoStop}%
\bibitem [{\citenamefont {Bortolussi}\ and\ \citenamefont
  {Palmieri}(2018)}]{bortolussi2018deep}%
  \BibitemOpen
  \bibfield  {author} {\bibinfo {author} {\bibfnamefont {L.}~\bibnamefont
  {Bortolussi}}\ and\ \bibinfo {author} {\bibfnamefont {L.}~\bibnamefont
  {Palmieri}},\ }\bibfield  {title} {\bibinfo {title} {Deep abstractions of
  chemical reaction networks},\ }in\ \href
  {https://link.springer.com/chapter/10.1007/978-3-319-99429-1_2} {\emph
  {\bibinfo {booktitle} {Computational Methods in Systems Biology}}}\ (\bibinfo
  {organization} {Springer},\ \bibinfo {year} {2018})\ pp.\ \bibinfo {pages}
  {21--38}\BibitemShut {NoStop}%
\bibitem [{\citenamefont {Thanh}\ and\ \citenamefont
  {Priami}(2015)}]{thanh2015simulation}%
  \BibitemOpen
  \bibfield  {author} {\bibinfo {author} {\bibfnamefont {V.~H.}\ \bibnamefont
  {Thanh}}\ and\ \bibinfo {author} {\bibfnamefont {C.}~\bibnamefont {Priami}},\
  }\bibfield  {title} {\bibinfo {title} {Simulation of biochemical reactions
  with time-dependent rates by the rejection-based algorithm},\ }\href
  {https://aip.scitation.org/doi/10.1063/1.4927916} {\bibfield  {journal}
  {\bibinfo  {journal} {J. Chem. Phys.}\ }\textbf {\bibinfo {volume} {143}},\
  \bibinfo {pages} {054104} (\bibinfo {year} {2015})}\BibitemShut {NoStop}%
\bibitem [{\citenamefont {Germain}\ \emph {et~al.}(2015)\citenamefont
  {Germain}, \citenamefont {Gregor}, \citenamefont {Murray},\ and\
  \citenamefont {Larochelle}}]{MathieuGermain2015MADEMA}%
  \BibitemOpen
  \bibfield  {author} {\bibinfo {author} {\bibfnamefont {M.}~\bibnamefont
  {Germain}}, \bibinfo {author} {\bibfnamefont {K.}~\bibnamefont {Gregor}},
  \bibinfo {author} {\bibfnamefont {I.}~\bibnamefont {Murray}},\ and\ \bibinfo
  {author} {\bibfnamefont {H.}~\bibnamefont {Larochelle}},\ }\bibfield  {title}
  {\bibinfo {title} {Made: Masked autoencoder for distribution estimation},\
  }\href {https://arxiv.org/abs/1502.03509} {\bibfield  {journal} {\bibinfo
  {journal} {arXiv:1502.03509}\ } (\bibinfo {year} {2015})}\BibitemShut
  {NoStop}%
\bibitem [{\citenamefont {Van~Oord}\ \emph {et~al.}(2016)\citenamefont
  {Van~Oord}, \citenamefont {Kalchbrenner},\ and\ \citenamefont
  {Kavukcuoglu}}]{van2016pixel}%
  \BibitemOpen
  \bibfield  {author} {\bibinfo {author} {\bibfnamefont {A.}~\bibnamefont
  {Van~Oord}}, \bibinfo {author} {\bibfnamefont {N.}~\bibnamefont
  {Kalchbrenner}},\ and\ \bibinfo {author} {\bibfnamefont {K.}~\bibnamefont
  {Kavukcuoglu}},\ }\bibfield  {title} {\bibinfo {title} {Pixel recurrent
  neural networks},\ }in\ \href {https://proceedings.mlr.press/v48/oord16.html}
  {\emph {\bibinfo {booktitle} {International conference on machine
  learning}}}\ (\bibinfo {organization} {PMLR},\ \bibinfo {year} {2016})\ pp.\
  \bibinfo {pages} {1747--1756}\BibitemShut {NoStop}%
\bibitem [{\citenamefont {Williams}(1992)}]{williams1992simple}%
  \BibitemOpen
  \bibfield  {author} {\bibinfo {author} {\bibfnamefont {R.~J.}\ \bibnamefont
  {Williams}},\ }\bibfield  {title} {\bibinfo {title} {Simple statistical
  gradient-following algorithms for connectionist reinforcement learning},\
  }\href {https://link.springer.com/article/10.1007%2FBF00992696} {\bibfield
  {journal} {\bibinfo  {journal} {Machine Learning}\ }\textbf {\bibinfo
  {volume} {8}},\ \bibinfo {pages} {229} (\bibinfo {year} {1992})}\BibitemShut
  {NoStop}%
\bibitem [{\citenamefont {Gardner}\ \emph {et~al.}(2000)\citenamefont
  {Gardner}, \citenamefont {Cantor},\ and\ \citenamefont
  {Collins}}]{gardner2000construction}%
  \BibitemOpen
  \bibfield  {author} {\bibinfo {author} {\bibfnamefont {T.~S.}\ \bibnamefont
  {Gardner}}, \bibinfo {author} {\bibfnamefont {C.~R.}\ \bibnamefont
  {Cantor}},\ and\ \bibinfo {author} {\bibfnamefont {J.~J.}\ \bibnamefont
  {Collins}},\ }\bibfield  {title} {\bibinfo {title} {Construction of a genetic
  toggle switch in escherichia coli},\ }\href
  {http://www.nature.com/nature/journal/v403/n6767/full/403339a0.html}
  {\bibfield  {journal} {\bibinfo  {journal} {Nature}\ }\textbf {\bibinfo
  {volume} {403}},\ \bibinfo {pages} {339} (\bibinfo {year}
  {2000})}\BibitemShut {NoStop}%
\bibitem [{\citenamefont {Neal}(2001)}]{neal2001annealed}%
  \BibitemOpen
  \bibfield  {author} {\bibinfo {author} {\bibfnamefont {R.~M.}\ \bibnamefont
  {Neal}},\ }\bibfield  {title} {\bibinfo {title} {Annealed importance
  sampling},\ }\href
  {https://link.springer.com/article/10.1023/A:1008923215028} {\bibfield
  {journal} {\bibinfo  {journal} {Stat. Comput.}\ }\textbf {\bibinfo {volume}
  {11}},\ \bibinfo {pages} {125} (\bibinfo {year} {2001})}\BibitemShut
  {NoStop}%
\bibitem [{\citenamefont {Hibat-Allah}\ \emph {et~al.}(2021)\citenamefont
  {Hibat-Allah}, \citenamefont {Inack}, \citenamefont {Wiersema}, \citenamefont
  {Melko},\ and\ \citenamefont {Carrasquilla}}]{hibat2021variational}%
  \BibitemOpen
  \bibfield  {author} {\bibinfo {author} {\bibfnamefont {M.}~\bibnamefont
  {Hibat-Allah}}, \bibinfo {author} {\bibfnamefont {E.~M.}\ \bibnamefont
  {Inack}}, \bibinfo {author} {\bibfnamefont {R.}~\bibnamefont {Wiersema}},
  \bibinfo {author} {\bibfnamefont {R.~G.}\ \bibnamefont {Melko}},\ and\
  \bibinfo {author} {\bibfnamefont {J.}~\bibnamefont {Carrasquilla}},\
  }\bibfield  {title} {\bibinfo {title} {Variational neural annealing},\ }\href
  {https://www.nature.com/articles/s42256-021-00401-3} {\bibfield  {journal}
  {\bibinfo  {journal} {Nat. Mach. Intell.}\ }\textbf {\bibinfo {volume} {3}},\
  \bibinfo {pages} {952} (\bibinfo {year} {2021})}\BibitemShut {NoStop}%
\bibitem [{\citenamefont {Tang}\ \emph {et~al.}(2022)\citenamefont {Tang},
  \citenamefont {Liu}, \citenamefont {Zhang},\ and\ \citenamefont
  {Zhang}}]{tang2022solving}%
  \BibitemOpen
  \bibfield  {author} {\bibinfo {author} {\bibfnamefont {Y.}~\bibnamefont
  {Tang}}, \bibinfo {author} {\bibfnamefont {J.}~\bibnamefont {Liu}}, \bibinfo
  {author} {\bibfnamefont {J.}~\bibnamefont {Zhang}},\ and\ \bibinfo {author}
  {\bibfnamefont {P.}~\bibnamefont {Zhang}},\ }\bibfield  {title} {\bibinfo
  {title} {Solving nonequilibrium statistical mechanics by evolving
  autoregressive neural networks},\ }\href {https://arxiv.org/abs/2208.08266}
  {\bibfield  {journal} {\bibinfo  {journal} {arXiv:2208.08266}\ } (\bibinfo
  {year} {2022})}\BibitemShut {NoStop}%
\bibitem [{\citenamefont {Cao}\ and\ \citenamefont
  {Liang}(2008)}]{cao2008optimal}%
  \BibitemOpen
  \bibfield  {author} {\bibinfo {author} {\bibfnamefont {Y.}~\bibnamefont
  {Cao}}\ and\ \bibinfo {author} {\bibfnamefont {J.}~\bibnamefont {Liang}},\
  }\bibfield  {title} {\bibinfo {title} {Optimal enumeration of state space of
  finitely buffered stochastic molecular networks and exact computation of
  steady state landscape probability},\ }\href
  {https://bmcsystbiol.biomedcentral.com/articles/10.1186/1752-0509-2-30#citeas}
  {\bibfield  {journal} {\bibinfo  {journal} {BMC Syst. Biol.}\ }\textbf
  {\bibinfo {volume} {2}},\ \bibinfo {pages} {1} (\bibinfo {year}
  {2008})}\BibitemShut {NoStop}%
\bibitem [{\citenamefont {Causer}\ \emph {et~al.}(2022)\citenamefont {Causer},
  \citenamefont {Ba\~nuls},\ and\ \citenamefont {Garrahan}}]{causer2021finite}%
  \BibitemOpen
  \bibfield  {author} {\bibinfo {author} {\bibfnamefont {L.}~\bibnamefont
  {Causer}}, \bibinfo {author} {\bibfnamefont {M.~C.}\ \bibnamefont
  {Ba\~nuls}},\ and\ \bibinfo {author} {\bibfnamefont {J.~P.}\ \bibnamefont
  {Garrahan}},\ }\bibfield  {title} {\bibinfo {title} {Finite time large
  deviations via matrix product states},\ }\href
  {https://doi.org/10.1103/PhysRevLett.128.090605} {\bibfield  {journal}
  {\bibinfo  {journal} {Phys. Rev. Lett.}\ }\textbf {\bibinfo {volume} {128}},\
  \bibinfo {pages} {090605} (\bibinfo {year} {2022})}\BibitemShut {NoStop}%
\bibitem [{\citenamefont {Cho}\ \emph {et~al.}(2014)\citenamefont {Cho},
  \citenamefont {Van~Merri{\"e}nboer}, \citenamefont {Bahdanau},\ and\
  \citenamefont {Bengio}}]{cho2014properties}%
  \BibitemOpen
  \bibfield  {author} {\bibinfo {author} {\bibfnamefont {K.}~\bibnamefont
  {Cho}}, \bibinfo {author} {\bibfnamefont {B.}~\bibnamefont
  {Van~Merri{\"e}nboer}}, \bibinfo {author} {\bibfnamefont {D.}~\bibnamefont
  {Bahdanau}},\ and\ \bibinfo {author} {\bibfnamefont {Y.}~\bibnamefont
  {Bengio}},\ }\bibfield  {title} {\bibinfo {title} {On the properties of
  neural machine translation: Encoder-decoder approaches},\ }\href
  {https://arxiv.org/abs/1409.1259} {\bibfield  {journal} {\bibinfo  {journal}
  {arXiv:1409.1259}\ } (\bibinfo {year} {2014})}\BibitemShut {NoStop}%
\bibitem [{\citenamefont {Kingma}\ and\ \citenamefont
  {Ba}(2014)}]{kingma2014adam}%
  \BibitemOpen
  \bibfield  {author} {\bibinfo {author} {\bibfnamefont {D.~P.}\ \bibnamefont
  {Kingma}}\ and\ \bibinfo {author} {\bibfnamefont {J.}~\bibnamefont {Ba}},\
  }\bibfield  {title} {\bibinfo {title} {Adam: A method for stochastic
  optimization},\ }\href {https://arxiv.org/abs/1412.6980} {\bibfield
  {journal} {\bibinfo  {journal} {arXiv:1412.6980}\ } (\bibinfo {year}
  {2014})}\BibitemShut {NoStop}%
\bibitem [{\citenamefont {Acharya}\ \emph {et~al.}(2022)\citenamefont
  {Acharya}, \citenamefont {Rudolph}, \citenamefont {Chen}, \citenamefont
  {Miller},\ and\ \citenamefont {Perdomo-Ortiz}}]{acharya2022qubit}%
  \BibitemOpen
  \bibfield  {author} {\bibinfo {author} {\bibfnamefont {A.}~\bibnamefont
  {Acharya}}, \bibinfo {author} {\bibfnamefont {M.}~\bibnamefont {Rudolph}},
  \bibinfo {author} {\bibfnamefont {J.}~\bibnamefont {Chen}}, \bibinfo {author}
  {\bibfnamefont {J.}~\bibnamefont {Miller}},\ and\ \bibinfo {author}
  {\bibfnamefont {A.}~\bibnamefont {Perdomo-Ortiz}},\ }\bibfield  {title}
  {\bibinfo {title} {Qubit seriation: Improving data-model alignment using
  spectral ordering},\ }\href {https://arxiv.org/abs/2211.15978} {\bibfield
  {journal} {\bibinfo  {journal} {arXiv:2211.15978}\ } (\bibinfo {year}
  {2022})}\BibitemShut {NoStop}%
\bibitem [{\citenamefont {Suzuki}(1976)}]{suzuki1976generalized}%
  \BibitemOpen
  \bibfield  {author} {\bibinfo {author} {\bibfnamefont {M.}~\bibnamefont
  {Suzuki}},\ }\bibfield  {title} {\bibinfo {title} {Generalized trotter's
  formula and systematic approximants of exponential operators and inner
  derivations with applications to many-body problems},\ }\href
  {https://link.springer.com/article/10.1007/BF01609348} {\bibfield  {journal}
  {\bibinfo  {journal} {Commun. Math. Phys.}\ }\textbf {\bibinfo {volume}
  {51}},\ \bibinfo {pages} {183} (\bibinfo {year} {1976})}\BibitemShut
  {NoStop}%
\bibitem [{Cod()}]{Code}%
  \BibitemOpen
  \href@noop {} {\emph {\bibinfo {title} {NNCME (GitHub, 2023);
  https://github.com/jamestang23/NNCME}}}\BibitemShut {NoStop}%
\end{thebibliography}
\end{document}